\documentclass[aps,pre,floatfix,twocolumn,superscriptaddress]{revtex4}
\usepackage{amsmath,amssymb}
\usepackage[usenames]{color}
\usepackage{graphicx}
\usepackage{epstopdf}
\usepackage{xspace}
\DeclareGraphicsExtensions{.eps}
\usepackage{mathbbol}

\newcommand{\be}{\begin{equation}}
\newcommand{\ee}{\end{equation}}
\newcommand{\bea}{\begin{eqnarray}}
\newcommand{\eea}{\end{eqnarray}}

\usepackage{rotating, graphicx}

\begin{document}

\author{Wanzhou Zhang}

\affiliation{College of Physics and Optoelectronics, Taiyuan University of Technology Shanxi 030024, China}

\affiliation{C.N. Yang Institute for Theoretical Physics and Department of Physics and Astronomy,
State University of New York at Stony Brook, Stony Brook, NY 11794-3840, USA}

\title{ Machine learning of phase transitions in the percolation and XY models}
\author{Jiayu Liu}

\affiliation{College of Physics and Optoelectronics, Taiyuan University of Technology Shanxi 030024, China}

\author{Tzu-Chieh Wei}
\affiliation{C.N. Yang Institute for Theoretical Physics and Department of Physics and Astronomy,
State University of New York at Stony Brook, Stony Brook, NY 11794-3840, USA}

\date{\today}
\begin{abstract}
In this paper, we apply machine learning methods to study phase transitions  in certain statistical mechanical models  on the two dimensional lattices, whose transitions involve non-local or topological properties,
including site and  bond percolations, the  XY model and the generalized  XY model.
We find that using just one hidden layer in a fully-connected neural network, the percolation transition can be learned
and the data collapse by using the average output layer gives correct estimate of the critical exponent $\nu$.
We also study the  Berezinskii-Kosterlitz-Thouless transition, which
involves binding and unbinding of topological defects---vortices and anti-vortices,  in the classical XY model.
The generalized XY model contains  richer phases, such as  the nematic phase, the paramagnetic and the quasi-long-range ferromagnetic phases,   and we also apply machine learning method to it.  We obtain a consistent phase diagram from the network trained with only data along the temperature axis at two particular parameter  $\Delta$ values, where $\Delta$  is the relative weight of pure XY coupling.
Besides using the spin configurations (either angles or spin components) as the input information in a convolutional neural network, we devise a feature engineering approach using the histograms of the spin orientations in order to train the network to learn the three phases in the generalized XY model and demonstrate that it indeed works. The trained
network by  using system size $L\times L$ can  be used to the phase diagram  for other sizes
($L'\times L'$, where $L'\ne L$) without any further training.

\end{abstract}
\maketitle

\section{Introduction}
Recent advancement in computer science and technology has enabled
processing  big data and artificial intelligence. Machine learning (ML) has been
 successfully and increasingly applied to everyday life,  such as digital recognition, computer
 vision,  news feeds, and even autonomous vehicles~\cite{Science}.  Besides that, ML methods have
also been recently adopted to various fields of science and engineering, and,  in particular,
in the context of
 phases of matter and phase transitions in physics~\cite{juan1,un1,un2,un3}.
The main tools are roughly divided into 	(I) supervised machine learning (classification or regression with labeled
training data) and (II)
unsupervised machine learning (clustering of unlabeled data)~\cite{mlbook}.

Amongst the earliest development in this direction of phases of matter,   it was used in the study of the thermodynamical  phase
transitions of the  classical Ising model and its gauge variant by supervised machine learning methods~\cite{juan1}. In addition,  unsupervised learning was also applied to
the Ising model and  the XY model mainly by the principal component analysis
 (PCA) method and  the autoencoder (an artificial neural network)~\cite{un1,un2,un3,un4}.  Other unsupervised  methods, such as  random trees embedding and t-distributed stochastic neighboring ensemble, have also been used; see e.g.~\cite{Chng}.
Instead of just learning the transition, a  learning scheme called confusion method was invented to  predict   the  phase transitions~\cite{confu1}, and similarly a  moving-window method  was also shown to be useful~\cite{confu2}.
Beyond classical physics, the quantum many-body problems have also been  studied with
artificial neural networks in the description of equilibrium and dynamical properties~\cite{qmbtroyer}. For example, the
strongly correlated Fermi systems were studied using e.g.   connected
networks~\cite{fm1},  self-learning methods~\cite{slfm2}, and even with ML methods beyond
limitation of the sign problems~\cite{bfm3}. Other systems  have also been studied successfully, such as
topological phases~\cite{tp1, tp2, tp3,tp4,tp5}, disorder systems~\cite{mbl}, quantum percolation model~\cite{qpm},
non-equilibrium models~\cite{nonequi,nonequi2} and
 many others~\cite{tomography1,pure,reinf,longising,Potts, impurity, boltzman,dnn}.
The machine learning has also been discussed in the context of tensor networks~\cite{tensor0,tensor1}, and  is helpful to
accelerate the Monte Carlo sampling and reduce the auto correlation time~\cite{m1,m2,m3}.
Attempts have been made to understand  theoretically by mapping it  to the renormalization group~\cite{normal0}.

Here we focus on using mostly supervised machine learning methods to study  two types of classical statistical models, whose transitions involve non-local or topological properties,
including site and  bond percolations, the  XY model and the generalized XY (GXY) models.
In order to apply supervised learning to calculate phase boundaries,
 one needs to prepare Monte Carlo configurations in three regimes~\cite{juan1}: (i)
below the suspected $T_c$, (ii) above the suspected $T_c$ and (iii)  an intermediate
regime in a region  containing $T_c$. The first two regimes are used
for training, by applying the trained algorithm to configurations in
the vicinity of $T_c$, one can infer an accurate $T_c$. However, if the purpose is to learn (instead of predicting) the phase transition, configurations from all three regimes are used for both training and testing, with the latter being  used to verify that the network indeed can learn the transition and the distinct phases with high confidence~\cite{juan1}.

When the model under study has a local
order parameter (OP) such as the magnetization,  the optimized fully-connected network (FCN)  actually can recognize   phase transitions by essentially averaging over
 local spins~\cite{juan1}. For the phase transition characterized by   the non-local order parameters, such as the topological phase of   the Ising gauge model~\cite{juan1} or the classical XY model~\cite{xycnn},  the
convolutional neural network (CNN) is a better tool than the FCN as it encodes spatial information. It was demonstrated in the classical 2d Ising gauge model that the optimized  CNN essentially uses violation of local energetic constraints to distinguish the low-temperature from the high-temperature phase~\cite{juan1}.

In percolation, non-local information such as the wrapping or spanning of a cluster is needed to characterize the phases and the transition in between.
Percolation is one of the simplest statistical physical models that exhibits a continuous phase transition~\cite{perco1,perco2,perco3}, and the system is characterized by a single   parameter,
the occupation probability $p$ of a site or bond, instead of temperature.
If a spanning cluster exists  in a randomly occupied lattice, then the configuration percolates~\cite{wrapping}.
Can the neural network be trained to recognize such nonlocal information and learn or even predict the correct critical point? If so, can it be used to reveal other properties of the continuous transition, such as any  of the critical exponents? This motivates us to study percolation using machine learning methods.

 We first use the unsupervised t-SNE method to characterize configurations randomly generated for various occupation probability $p$ and find that it gives clear separation of configurations away from the percolation threshold $p_c$. This indicates that other machine learning methods such as supervised ones will likely work, which we also employ.
We find that both the FCN and the CNN works for learning the percolation transition, and the networks trained with configurations labeled with information of whether they are generated with $p>p_c$ or $p<p_c$ can result in a data collapse for different system sizes, giving the critical exponent of the correlation length.  We alternatively train the neural  network to learn the existence of a spanning cluster and with this the percolation transition can be identified (without supplying labels of $p>p_c$ or $p<p_c$).

Our interest in the XY model originates from its topological properties---vortices and how ML methods can be used to learn the BKT transition. The XY model in the two-  and three-dimensional lattices  were studied  by the unsupervised PCA method~\cite{un2,un3} and generative model~\cite{tow}. In Ref.~\cite{un3},
various choices of input were considered, e.g.  spin configurations (i.e. components of spins), local vortices, and their square into the PCA method. It was concluded that learning the vortex-antivortex unbinding to predict the transition might be difficult in the PCA.
 In a very recent work by Beach, Golubeva and Melko~\cite{xycnn}, it was shown that the  CNN works better than FCN  to learn the BKT transition. Furthermore, the advantage of using vorticies rather than spin configurations only shows upfor large system sizes, e.g. $L\gtrsim32$, but for smaller sizes using spin configurations may work better. Here, for small sizes we use either spin orientations or their components as input to a CNN and verify that both give successful learning of the BKT transition in the XY model. Additionally, we find that using the histograms of the spin orientations also works for the XY model and can be applied efficiently to larger system sizes Additionally, we find that using the histograms of the spin orientations also works for the XY model and the training can be done efficiently for larger system sizes.
To go beyond the XY model, we find it interesting to apply the ML to the generalized  XY (GXY) model~\cite{xyq2,xyq3} as it contains more complex configurations such as  half-vortices linked by strings (domain walls) and an additional nematic phase.
We find the use of spin configurations (either angles or spin components) and histograms both works.
The advantage of the latter approach  is that the training can be done for larger system sizes and, moreover,
trained network for one system size can be applied to other system sizes.

\begin{figure}[tbh]
\centering
\includegraphics[width=0.48\textwidth]{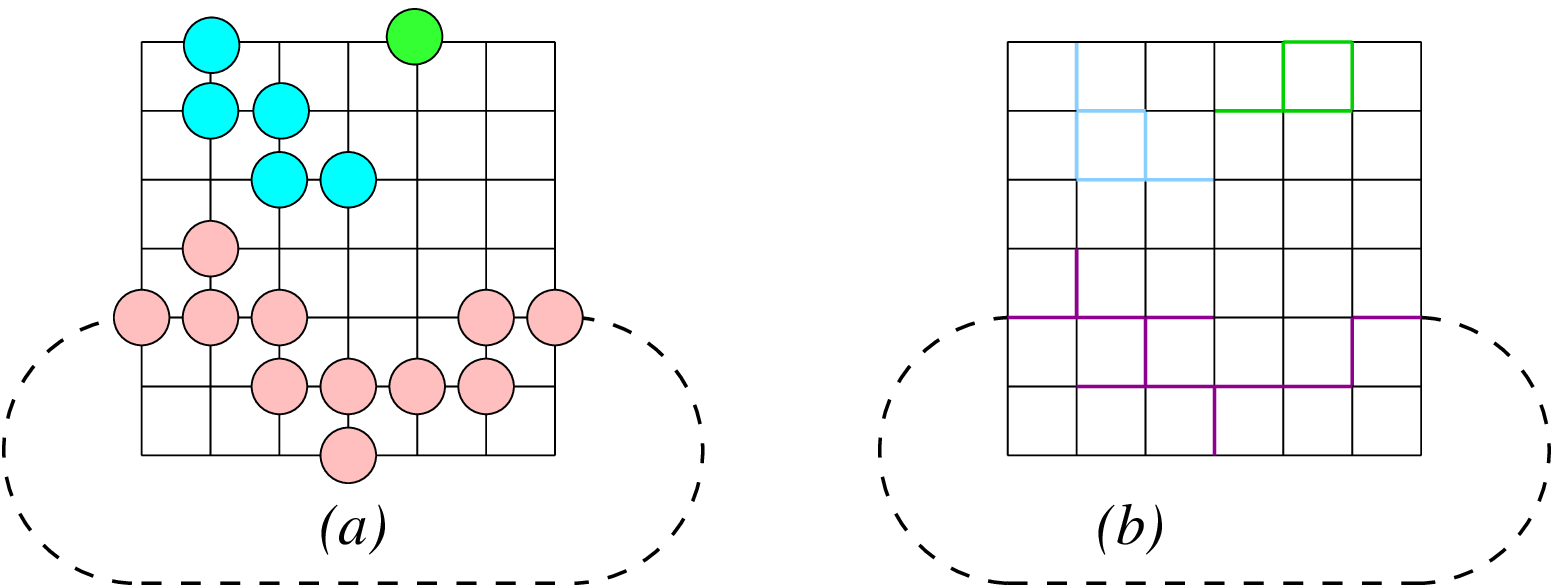}
\vspace{0.2cm}
\includegraphics[width=0.43\textwidth]{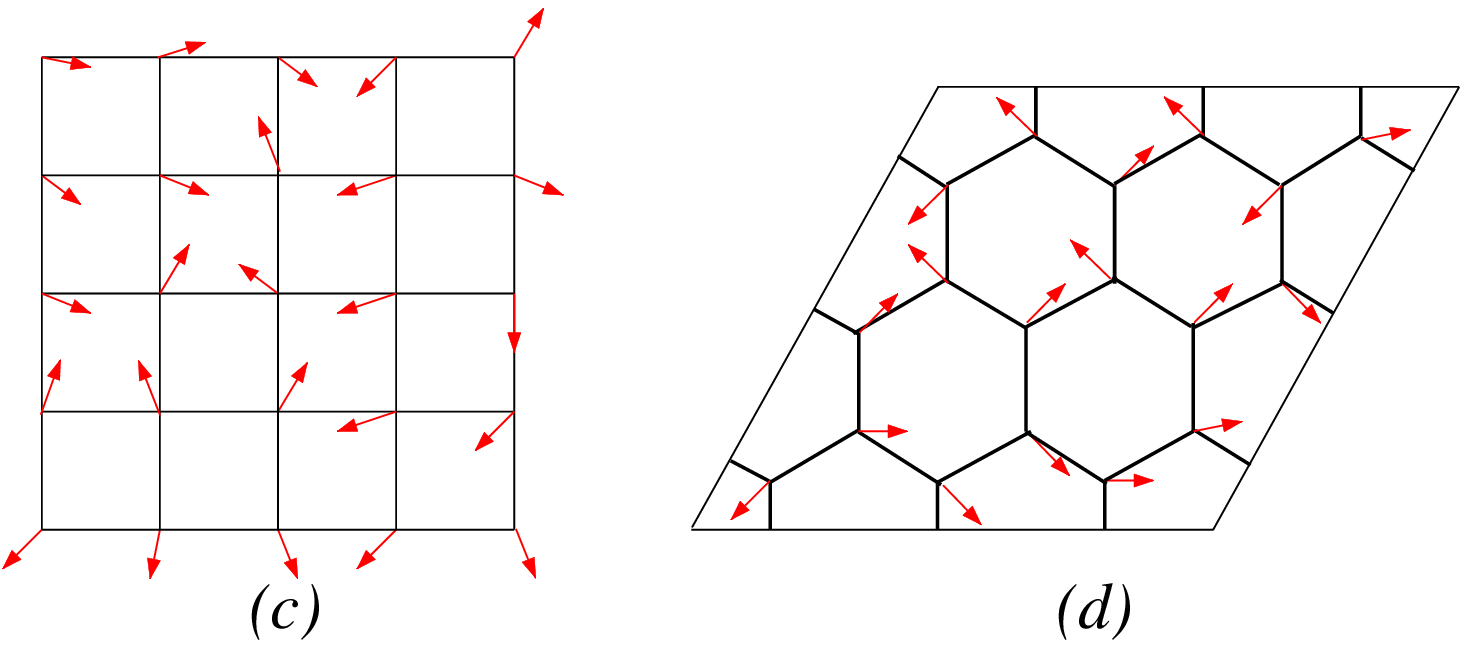}
\caption{(Color online)
 (a) A site percolation configuration on the square lattice with periodic boundary condition. With the periodic boundary condition,  the dashed line connects the opposite sides of the largest cluster  and it demonstrates the ``wrapping". 	 (b) A bond percolation configuration    on a square lattice.
 (c) Configuration of the XY model on the square lattice.  (d) Configuration of the XY model on the honeycomb lattice. }
\label{fig:x-net}
\end{figure}

The outline of this work is as follows. In Sec.~\ref{sec:model}, we  introduce models to be studied, i.e.,
site and bond percolations, the XY model and the GXY models.
In Sec.~\ref{sec:tsne}, we show classifications via the t-SNE approach on the high-dimensional configurations of site and bond percolation on both square and triangular  lattices. Then supervised learning using the fully-connected neural network is illustrated in Sec.~\ref{sec:fcn}.
 In  Sec.~\ref{sec:cnns},  the
CNN structure is introduced and the learning results of
percolations  by using the CNN are described, and a different  way of labeling  the configuration (using the existence of a spanning cluster) is  used. No labeling regarding $p>p_c$ or $p<p_c$ is used there, but the transition point can be obtained.
In Sec.~\ref{sec:XYmodels} we use the CNN to study the  XY model and  its generalized versions.
For the GXY model containing  a nematic phase, we construct the histograms of the spin orientations and then  use them as images for the network to learn.
This way of feature engineering can result in better learning of the transitions. We conclude in Sec.~\ref{sec:con}.

\section{Models}
\label{sec:model}
In this paper we study two types of classical statistical physical models, include (I) site and  bond percolations, and (II) the  XY model and (III) the GXY models. Let us introduce them as follows.

\smallskip\noindent (Ia)
{\it  Site percolation}.
The site percolation can be defined by the  partition function,
\begin{equation}
     Z=\sum_{\{\sigma\}}p^{n_s^{\sigma}}_s(1-p_s)^{N-n_s^{\sigma}},
\end{equation}
where, e.g., on the square lattice  with $N=L\times L$  sites, $p_s$ is the probability
of site occupation, $n_s^{\sigma}$ is the numbers of sites being occupied in the configuration labeled by $\sigma$.
In order to  obtain the critical  phase transition points by Monte-Carlo simulations, usually the wrapping  probability $R$
is defined~\cite{wrapping} in the case of the periodic boundary condition.
With open boundaries, a cluster growing large enough could touch the two
opposite boundaries and hence it is referred to a spanning cluster.
The wrapping cluster is defined as a cluster that connects
opposite sides that would be in the otherwise open boundary condition, as illustrated in Fig.~\ref{fig:x-net} (a) and (b).
A cluster forming along either the $x$ or $y$ direction can contribute to  $R$.
In ML  method, we do not need to measure this observable directly and a naive labeling of  each configuration is given according to how it is generated according to the occupation probability $p$ and $p$'s relation with the critical value (the percolation threshold) $p_c$, i.e., whether $p>p_c$ (say labeled as `1') or  $p<p_c$ (say labeled as `0'). This resembles the scenario in the Ising model whether configurations are labeled according to whether they are generated above or below the transition temperature $T_c$.
But such a topological property of wrapping (or percolating) can be used as an alternative labeling for training the neural network (see Sec.~\ref{sec:lci}).

\medskip\noindent (Ib) {\it Bond percolation}.
The bond percolation partition function can be defined as,
\begin{equation}
     Z=\sum_{\{\sigma\}}p_b^{n_b^{\sigma}}(1-p_b)^{N-n_b^{\sigma}},
\end{equation}
where $p_b$ is the probability of occupying a bond on the lattices, $n_b$ is the number of bonds being occupied in the  bond configuration $\sigma$. The wrapping or spanning cluster is defined in a similar way as in the site percolation.

\medskip\noindent (II) {\it XY model}.
 The Hamiltonian of the classical XY model~\cite{xypoints} is given by
\begin{equation}
  H = -J\sum_{\langle i,j\rangle}\vec{s_i}\cdot\vec{s_j} = -J\sum_{\langle i,j\rangle} \text{cos}(\theta_i-\theta_j),
\end{equation}
where $\vec{s_i}$ is unit vector with two real components and $\langle i,j\rangle$ denotes a nearest-neighbor pair of sites $i$  and $j$, and  $\theta_i$ in $(0,2\pi]$ is a classical
variable defined at each site. The sum in the Hamiltonian is over nearest-neighbor pairs or bonds on the square
lattice ($L \times L$) with the periodic boundary condition; other lattices can be also considered.

\medskip\noindent(III) {\it Generalized XY model}.
The Hamiltonian of the classical GXY models is given by
\be
\label{eqn:GXY}
H=-\sum [\Delta \text{ cos}( \theta_i - \theta_j)+ (1-\Delta) \text{cos} (q\theta_i-q \theta_j)],
\ee
where $\Delta$ is the  the relative weight of the pure XY model, and $q$ is another integer parameter that could drive  the  system to form a  nematic phase.
For both $\Delta=0$ and 1 the model reduces to the
pure XY model (redefining $q\theta$ as $\theta$ in the first case),
and hence the transition temperature is identical to that of the pure
XY model. However, such a redefinition is not possible with $\Delta\ne 1$.  The phase diagrams of the GXY models~(\ref{eqn:GXY}) depend on the integer parameter $q$, and they have been explored  extensively~\cite{xyq2,xyq3}.

\section{percolations}
\label{sec:percolation}
Even though we mainly use supervised learning methods, we will begin the study of percolation using an unsupervised method, the t-distributed stochastic neighbor embedding (t-SNE).  We shall see that it can characterize configurations randomly generated in percolation to two distinct groups with high and low probabilities $p$ of occupation as well as a belt containing configurations generated close to the percolation threshold $p=p_c$. This gives us confidence to proceed with supervised methods such as FCN and CNN.
\subsection{Learning percolation by t-SNE}
\label{sec:tsne}

\begin{figure}[htb]
\centering
\includegraphics[width=0.4\textwidth]{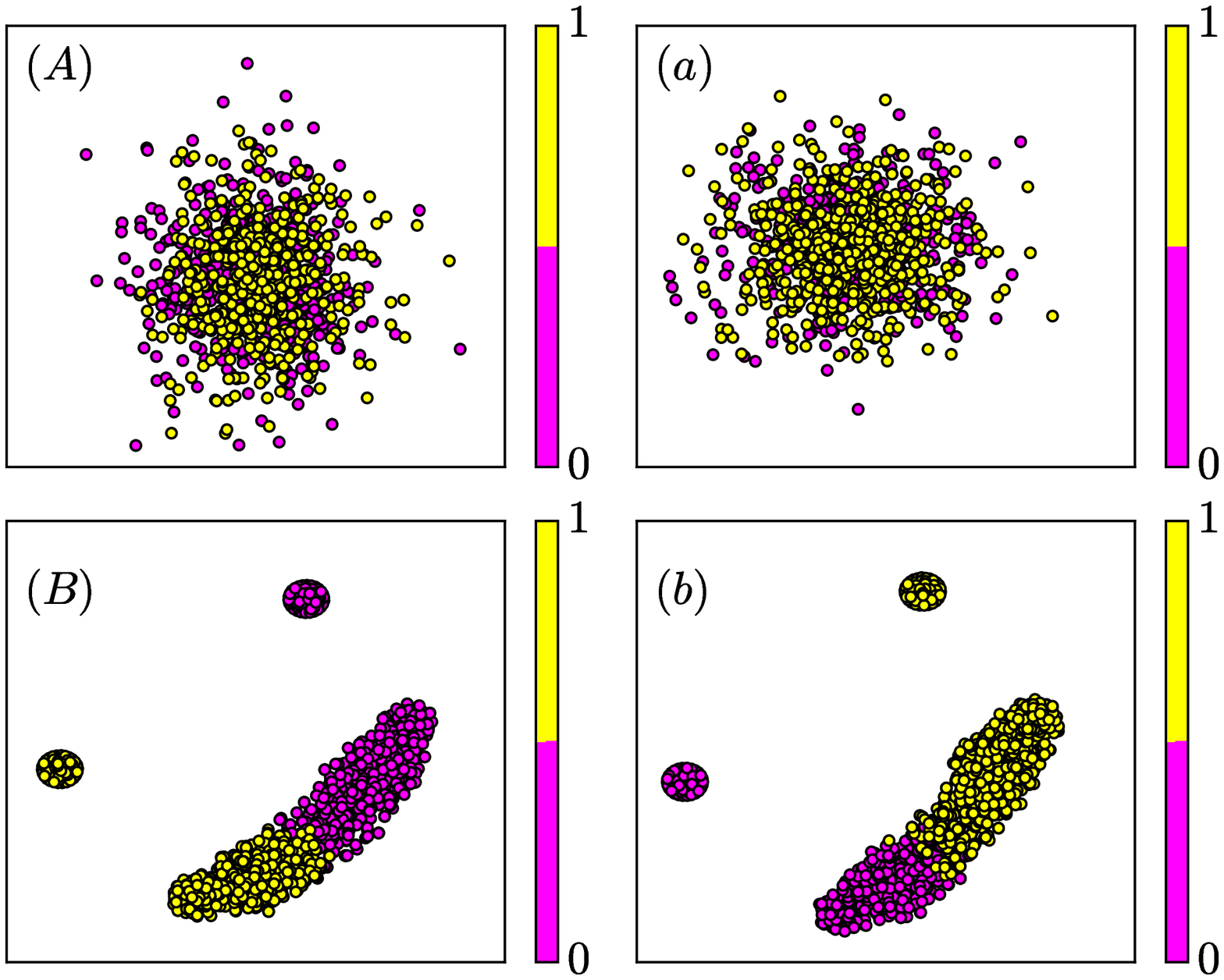}
\includegraphics[width= 0.4 \textwidth]{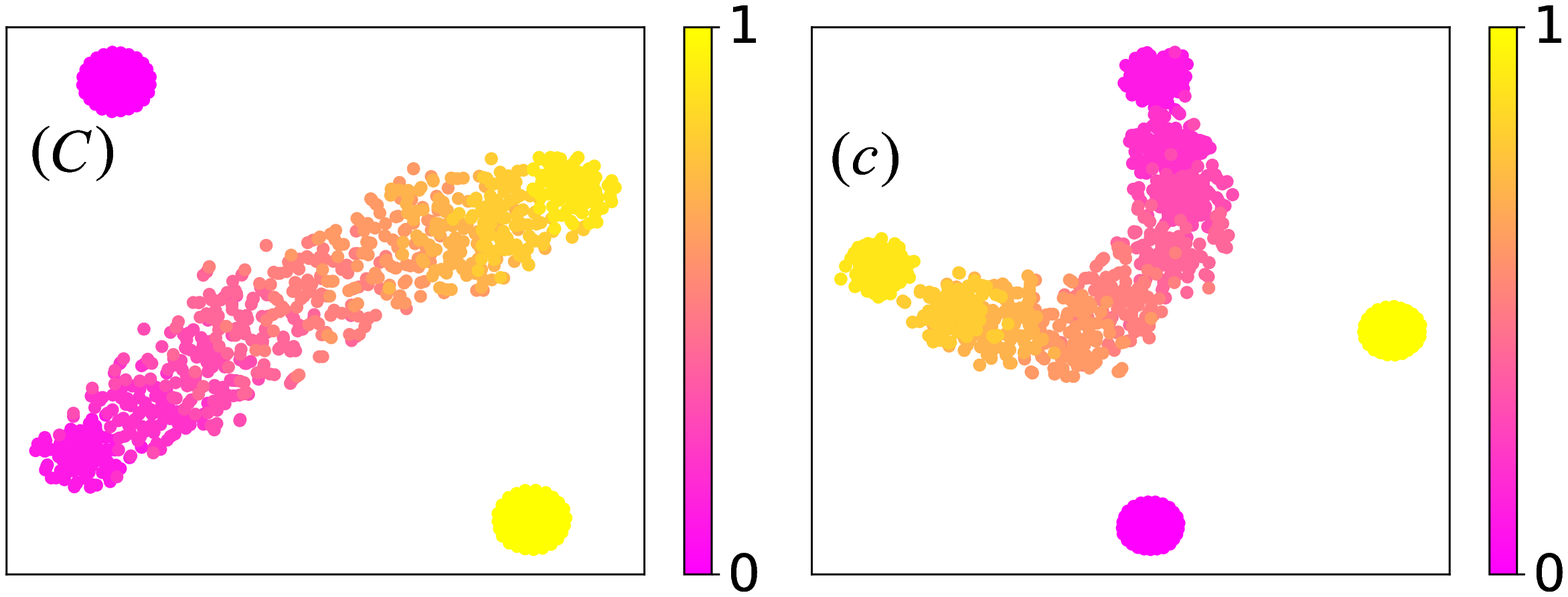}
\caption{(Color online)
The t-SNE distribution of site percolation on the square lattice with $L=32$ for
(A) one step of iteration,  and (B)  2000 steps of  iterations. (C) The result for bond percolation on the square lattice with size $L=16$ after  1100 iterations.
The t-SNE distribution  of	site percolation on the triangular  lattice with $L=32$ for
  (a) only  one step of iteration and (b)  2000 steps of iteration. (c) The result for the bond percolation on triangular lattice at $L=16$ after 1100 iterations. Note that in (A), (a), (B) and (b), we only use two colors pink and yellow, but in (C) and (c) the samples are colored according to the probability $p$ of occupation,  from pink to yellow.  }
\label{tsne}
\end{figure}
The t-SNE is an unsupervised machine learning
 algorithm for dimensionality reduction via an iteration procedure~\cite{tsne1}. By using
a nonlinear technique (unlike the PCA), it  projects high-dimensional data
 (e.g. $M$-dimensional objects $\boldsymbol{x}_1, \dots, \boldsymbol{x}_N$) into a two-dimensional space, which can
 then be visualized in a scatter plot, where similar (or dissimilar) objects are
 modeled by nearby (or distant) points. (Here, the bold symbols means that  each $\boldsymbol{x} $ is a $M$-component vector.)  For example, it has been successfully used to analyze the Ising configurations and
project the data into two-dimensional scattering figures~\cite{juan1}.

We show, in  Fig.~\ref{tsne} (A), for site percolation on the square lattice with size $L=32$, such a scatter plot, produced
  by using $M=11000$  site configurations in the t-SNE procedure, where each
  configuration $\boldsymbol{x}$ contains $32\times 32=1024$ elements 0 or 1. Fig.~\ref{tsne} (A)  is the distribution obtained after only one step of  iteration in the t-SNE method.
 Clearly, after the first iteration, the data for both labels are still mixed together and there is no separation into distinct groups.
However, after 2000 iterations, as shown in Fig.~\ref{tsne} (B), the data converges into three distinct groups, with two concentrated clusters and a wide `belt'. The  concentrated cluster with yellow solid  circles
  indicates data generated from non-percolating (or subcritical) phase, i.e. $p<p_c$, while  the purple  cluster indicates data from the percolating (or supercritical) phase, i.e. $p>p_c$. In addition to the two distinct clusters,
  the belt contains the data around the percolation transition point $p_c$ (roughly between 0.2 and 0.8).
Similar behavior  in the t-SNE analysis of the distribution
is also obtained in the percolation study on the triangular lattice, as shown in Figs.~\ref{tsne} (a) \& (b).

We remarked that, there are only two colors used in Figs.~\ref{tsne} (A), (a), (B) \& (b), indicating only above or below $p_c$, but a continuous hue
between purple and yellow was used in Figs.~\ref{tsne}(C) and (c) to denote
the occupation probability $p$.
In Fig.~\ref{tsne} (C) \& (c), we  show the embeddings for occupation probability $p$ of the {\it bond} percolation on
the square lattice  and on the triangular lattice, respectively, both with $L= 16$. The behavior is similar to that of site percolation.  The upshot is that the t-SNE method can characterize percolation configurations  into different phases and near the transition. In order to obtain the transition point $p_c$, one can probably divide the belt into two halves and the probability value $p$ at the cut can be used as an estimation of the percolation threshold. But we do not do that here, as we will use supervised learning below to learn the transition  more accurately.

\subsection{ Learning  percolation by FCN}
\label{sec:fcn}
\begin{figure}[htb]
\centering
\includegraphics[width= 0.48 \textwidth]{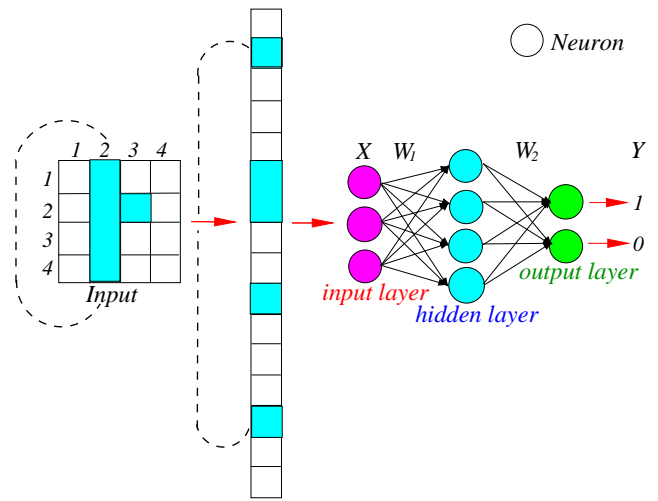}
\caption{(Color online)
Architecture of the fully connected neural network used to obtain $p_c$ for the site percolation model.
  The  input
is a set of =  configurations  in  a  two-dimensional
lattices with periodic lattice with size $L=4$. A  percolating configuration is illustrated, where
the blue squares are the occupied sites while the empty squares are not occupied. }
\label{fig:x-fcn}
\end{figure}

 In Fig.~\ref{fig:x-fcn}, we  show the structure of the FCN (which we implemented with the  TensorFlow library~\cite{tf}), which   consists  of one input layer, one hidden layer and
 one output layer of neurons. The network between two neighboring layers is fully-connected, i.e.,   each neuron in one layer is connected to every neuron in the previous and
 next layer.
The layers are interconnected by sets of correlation weights (usually denoted by a matrix $ \boldsymbol{W}$) and there are biases (denoted by a vector $\boldsymbol{b}$ associated with neurons at each layer (except the input one).
The input layer  accepts the   data sets of images or configurations  and then
 the network   processes the data according to the weights and biases, as well as some activation function for each neuron. The optimization is performed to minimize some cost function, e.g., the cross entropy function in Eq.~(\ref{eqn:CE})  via the stochastic gradient decent method.  {The number of  neurons in the input layer should be equal to
 the total number of lattice sites, i.e. $L\times L$ in the site percolation (or the number of spins in the Ising model), or the total number of bonds in the bond percolation}. The values of the neurons
are denoted as the input vector $\boldsymbol{x}$.

 The hidden layer  is to transform the inputs into something that the output layer can use and one can employ as many such hidden layers (but we will focus on just one hidden layer in FCN).
 It determines the mapping relationships. For example, as illustrated in Fig.~\ref{fig:x-fcn}, we denote  $\boldsymbol{W_1}$ as the weight matrix from the input to the hidden layer and $\boldsymbol{b_1}$ the bias vector for the hidden layer. {The number of neurons in the hidden layers generally is chosen to be approximately of the same order as the size of the input layer or less.}
  Assume the activation function for the neurons in the hidden layer is $f$. Then the neurons in the hidden layer will output $\boldsymbol{y_H}={f}(\boldsymbol{x}\cdot \boldsymbol{W_1} +\boldsymbol{b_1})$. Denote $\boldsymbol{W_2}$  as the weight matrix from the hidden to the output layer and $\boldsymbol{b_2}$ the bias vector for the output layer. Assume the activation function for the neurons in the hidden layer is ${g}$. Then the neurons in the output layer will have states described by $\boldsymbol{y}=g(\boldsymbol{y_H}\cdot\boldsymbol{W_2}+\boldsymbol{b_2})$.  One choice  usually used as the activation is the so-called {\it sigmoid} function,
\begin{equation}
\sigma(z) \equiv \frac{1}{1+e^{-z}},
\end{equation}
related to the Fermi-Dirac distribution in physics.
Another is the so-called {\it softmax} function that, when applied to a vector with components ${z}_j$, gives
another vector  with components.
\begin{equation}
a_j = \frac{e^{{z}_j}}{\sum_k e^{{z}_k}},
\end{equation}
and it is related to the Boltzmann weight in physics.
Yet another choice that has become popular recently is the rectifier, defined as $f_{\rm Rec}(z)=\max(0,z)$, and one unit that uses this activation function is called a rectified linear unit.  In principle, one can employ as many hidden layers in the network. It is the universality of the neural network, i.e., it can approximate any given function, that makes machine learning powerful.

 These weights $\boldsymbol{W}$'s and biases $\boldsymbol{b}$'s need to be optimized at the training stage. The inputs consist of pairs of $\{\boldsymbol{x}:\boldsymbol{y_T}\}$, where each configuration is represented by a vector $\boldsymbol{x}$ of $L\times L$ components of value being 1 (whether a site is occupied) or 0 (not occupied) and a corresponding label $\boldsymbol{y_T}$ indicating whether the configuration $\boldsymbol{x}$ is generated above or below the percolation threshold $p_c$.
This label can be described by by one single
binary number, e.g. 1 representing $p>p_c$ and  0 representing $p<p_c$. The cost function can be chosen as (i) the average two-norm between the label vectors $\boldsymbol{y_T}$ and the output layer vector  $\boldsymbol{y}$ (resulting from input $\boldsymbol{x}$) over many such pairs,
 \begin{equation}
 L_2\equiv\frac{1}{N}\sum_{\boldsymbol{x}} |\boldsymbol{y}(\boldsymbol{x})-\boldsymbol{y_T}(\boldsymbol{x})|^2,
 \end{equation}
 or (ii) the average cross-entropy between such pairs,
  \begin{equation}
  \label{eqn:CE}
C_E\equiv -\frac{1}{N}\sum_{\boldsymbol{x}}\sum_j \big(\boldsymbol{y_T}_j \log \boldsymbol{y}_{j}+(1-\boldsymbol{y_T}_j) \log(1- \boldsymbol{y}_{j})\big).
 \end{equation}
An additional term called regularization, such as $\lambda/(2N) \sum_{\boldsymbol{i}}|\boldsymbol{W_i}|^2$, is introduced to the cost function in order to prevent overfitting. The optimization is done with stochastic gradient descent using the TensorFlow library.

\begin{figure}[htb]
\centering
\includegraphics[width=0.48\textwidth]{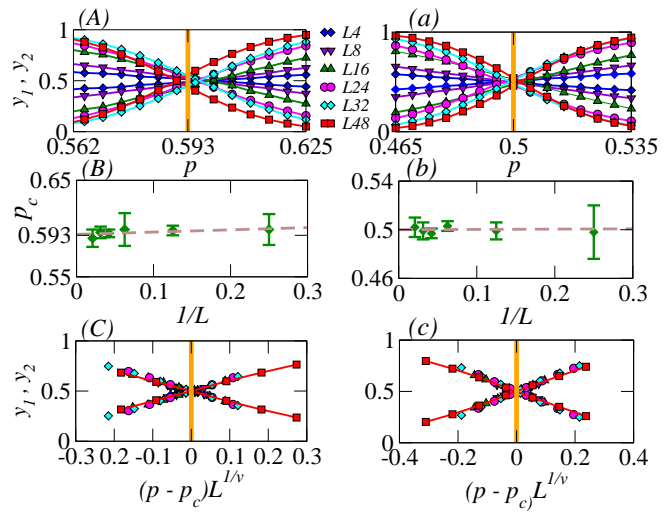}%
\caption{ {(Color online)  The two values $y_1$ and $y_2$ of the output layer of  site percolation on the square lattice (left columns) and triangular lattice (right columns)  with the system sizes $L=4, 8, 16, 24, 32, 48$ using
	the fully connected network. The  second and last  rows are, respectively, the average results of the output layer ($y_1$, $y_2$), and the finite size scaling for extracting the critical point, the data collapse of $y_1$ and $y_2$. }
}
\label{fig:fcn}
\end{figure}

Once the network is optimized after the training stage, we use as input  different and independently generated configurations with possibly different sets of $p$ values, and use the average values  of the output layer $\boldsymbol{y}$ (sometimes referred to as the average output layer) from the network to estimate the transition. This is referred to as the test stage.
The two components in $\boldsymbol{y}$ are in the range $[0,1]$, and the larger the value of the component (associated with a neuron) gives the more probable  prediction the neuron makes.
Usually we plot such average numbers for both output neurons (one is associated label 0 and the other 1).  This results  in two curves as a function of the probability $p$, as illustrated in Fig.~\ref{fig:fcn}. The value of $p$ at which the two curves cross is used as an estimate of the percolation threshold. Note that, as seen below in Sec.~\ref{sec:XYmodels}, when we encounter three or more phases to identify, then the number of neurons in the output layer will be accordingly three or more.

\vskip 0.5cm
\begin{figure}[ht]
\centering
\vskip 0.4 cm
\includegraphics[width=0.4\textwidth]{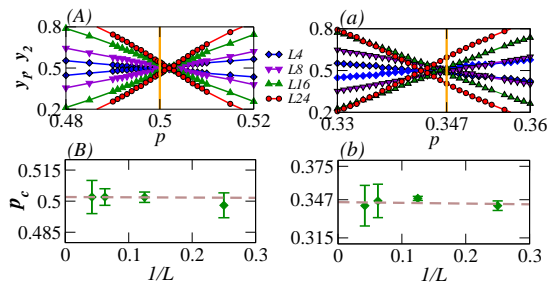}
\caption{{(Color online)  The two values $y_1$ and $y_2$ of the output layer of the machine learning for bond percolation on the square lattice (left columns) and triangular lattice (right columns)  with the system sizes $L=4, 8, 16, 24$ using   the fully connected network. The first and the second rows show the average results in the output layer $y_1$ and $y_2$ and the finite size scaling for the critical points, respectively. }}	\label{fig:fcn1} \end{figure}

In Fig.~\ref{fig:fcn} (A), we show the two average  values $y_1$ and $y_2$ in the output layer, obtained by testing the network with site percolation configurations {in the range
of $0.562< p < 0.625$ for different system sizes}. {The FCN was trained  using data
 from $0<p<1$ with 2500 samples of  per value of $p$ . To reduce the statistical errors, 20000 samples per $p$ are used  for the test data to obtain two neurons $y_1(p)$ and $y_2(p)$, i.e.
 the average values of the output layer.}   We find that the two corresponding lines cross, and the crossing  in Fig. ~\ref{fig:fcn}(B) as function of $1/L$ is used to
estimate the transition point in the thermodynamic limit $L\rightarrow \infty$. From this,
$p_c$ is estimated to be $0.594\pm0.002$.  This agrees with the phase transition from the Monte-Carlo methods within error bars~\cite{sqpc}.
In Fig. ~\ref{fig:fcn} (C), we show  the data collapse of the
two average output layers.
Due to the finite size effects, the intersects between  different sizes $L=4, 8, 16, 24, 32$ and $48$,  are slightly   shifted away from the exact $p_c$.
By taking into account of this and by rescaling  the horizontal  axis with $(p-p_c)L^{1/\nu}$,  the data  collapse very well to a single curve for each neuron output,  with the use of the exponent $\nu=4/3$ from  percolation theory~\cite{exp}.

During the stage of training, the labeling $\boldsymbol{y_T}$ gives the information whether   a configuration is generated at the probability $p$ greater or less than $p_c$.  The information about whether the individual configuration is percolating or not is not known. However, one would expect that for the configurations generated sufficiently away from $p=p_c$, the neural network is learning such a property. But close to $p_c$ even if a configuration is generated at $p<p_c$ it can still be percolating and vice versa even if a configuration is generated at $p>p_c$ it may not be percolating. There are a lot of fluctuations near $p_c$; see also Figs.~\ref{clabel} (e) and (f). In Sec.~\ref{sec:lci}, we will use the alternative labeling by giving the information of whether a configuration is percolating or not.

We also apply the FCN to site percolation on other lattices. For example, on the triangular lattice, the
 percolation threshold $p_c=0.5$ is also exactly known~\cite{bptri}.
Similar to the square lattices, we generate the configuration  with  randomly occupied  sites  with a statistically independent probability $p$ and then label the configurations according to whether $p>p_c$ or $p<p_c$. After training the FCN, we test FCN with different data sets, the average values of the output layer cross at
the phase transition point $p_c=0.5$~\cite{bptri}.
The data collapse and the finite size scaling are shown in Figs.~\ref{fig:fcn} ($b$) and ($c$), respectively.

Similarly, we use the FCN to study bond percolation, and the results are shown in
Figs.~\ref{fig:fcn1} (A) and (B) give  the average output layer for
the square lattice (with
sizes $L=8, 16, 24, 32$), and
finite size scaling (yielding the critical point at $p_c=0.5$), respectively.
    Figs.~\ref{fig:fcn1} (a) and (b) show the results of bond percolation on
  the triangular lattices. The bond percolation threshold~\cite{bptri} is at  $ 2 \sin (\pi/18)\approx0.347$
  and our result agrees with it.  We remark that in the training, each input configuration has
$2L\times L$ bond variables for the square lattice
and  $3 L\times L$ bond variables for triangular lattices.

\begin{figure}[tbh]
\includegraphics[width=0.35 \linewidth]{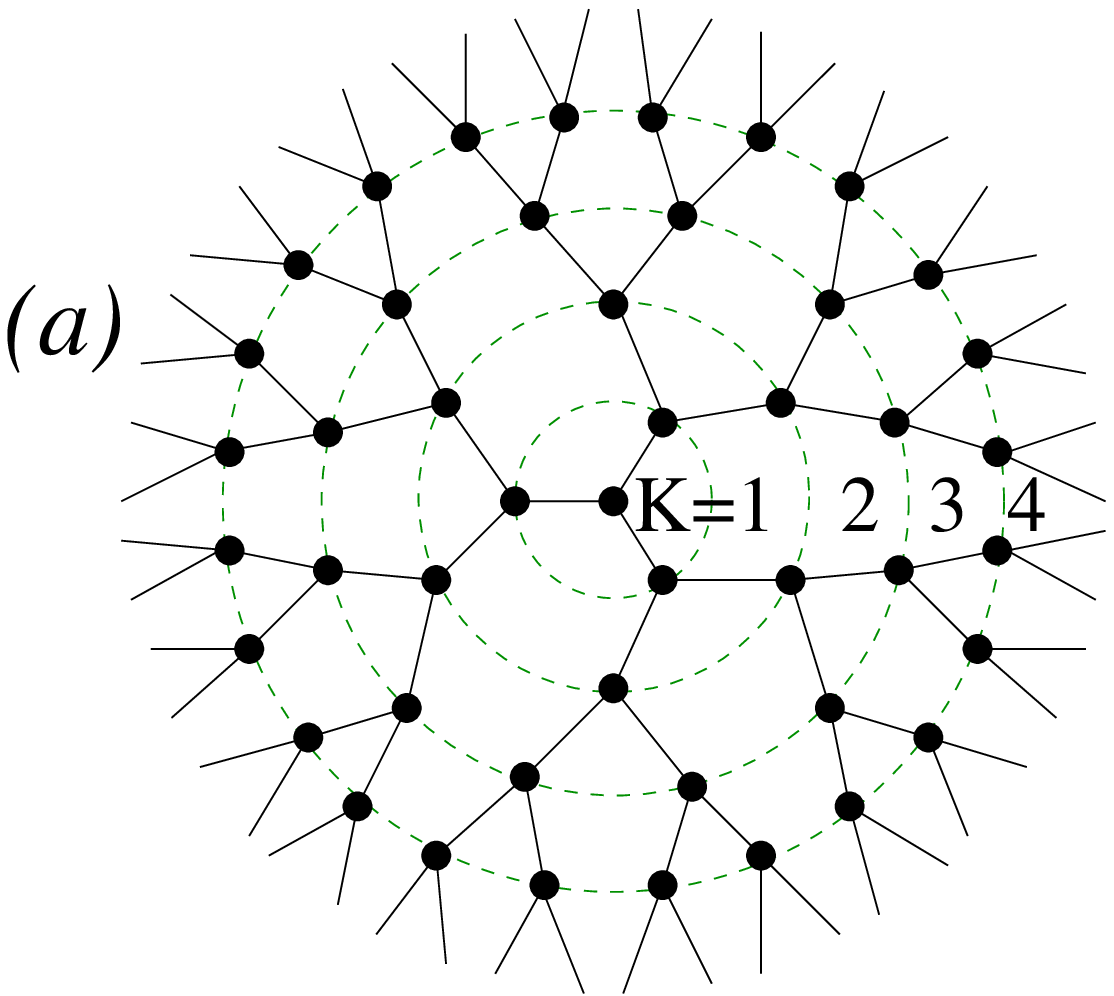}
\hskip 0.1 cm
\includegraphics[width=0.5 \linewidth] {Fig6b.eps}
\caption{
	(Color online)
	(a)  Bethe lattice with coordination number $z = 3$. The lattice sites are represented by solid circles at different shells $K = 0, 1, 2$, $\cdots$.  (b) The average values $y_1$ and $y_2$ in the output layer  of site percolation on the Bethe lattice for $K=3$ and $K=5$.}
\label{bethe}
\end{figure}

Site percolation on the Bethe lattice was studied analytically and exact transition was known, and thus it is interesting to apply the neural network. In Fig.~\ref{bethe} (a)  we illustrate the Bethe lattice with four shells, indicated by  the green dashed lines. Different from the square or triangular lattices,
the Bethe lattice with coordination number $z$ (here $z=3$) has  a topological  tree structure that expands from a central site out to infinity~\cite{bethe}.
Each site has  one neighboring site pointing towards the central site  and $z -1$ sites going away from it.
The total number of sites in $K$-th shell  is $ N_K = z(z-1)^{(K-1)} $.
 {Checking whether the configuration  percolates or not  by machine learning is interesting because the path connecting  any two sites of different trees has to go through the central site.}
The exact critical probability~\cite{bethe} is known to be  at $p_c=0.5$ and  our learning  results  in Fig.~\ref{bethe} (b)  show that  the FCN can recognize
the phase transition after training the network. The results are obtained using $K= 3$ and  5 and the total size is  $N=1+\sum_1^K N_K$, i.e.,  22 and  94.

\subsection{Learning percolation by CNN}
\label{sec:cnns}
We have seen in the previous section that the FCN works  well in learning percolation transition. However the information about the lattice structure is not explicitly used, but rather it might be inferred during optimization. For problems that have such natural  spatial structure, the CNN is naturally suited and can yield better results.
\subsubsection{CNN structure}

\begin{figure}[bh]
\centering
\includegraphics[width= 0.48 \textwidth]{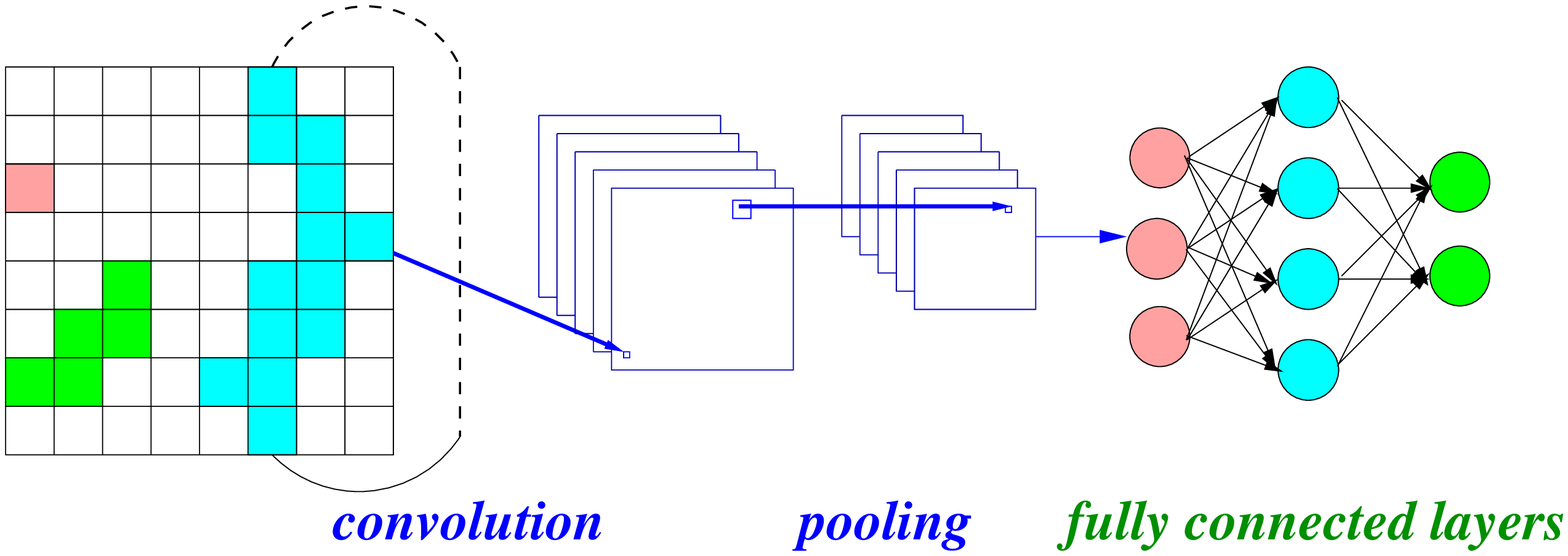}
\caption{(Color online)
Architecture of the fully connected neural network used to obtain $p_c$ for the site percolation model.
  The  input
is  the    configurations  in  a  two-dimensional
lattice with periodic boundary condition, illustrated  with size $L=4$. The
the colored or shaded squares are the occupied sites while the white squares are not occupied. }
\label{fig:cnn}
\end{figure}

We first begin by discussing the structure of the CNN shown in Fig.~\ref{fig:cnn}, where  the input is  a two-dimensional array or an image. There is a filter with a small size such as $5\times 5$ also  called a local receptive field, that processes information of a small region. This same local receptive field moves along the lattice to give a coarse-grained version of the original 2D array or image. We can move the filter not one lattice site but a few (which is usually called {\it stride\/} to the next region). We can also pad the outer regions with columns or rows of zeros so as to maintain the same size of the filtered array, which is referred to as {\it padding\/}.   This results in a filtered or generally coarse-grained hidden layer. We can use many different local receptive fields to obtain many such layers, usually referred to as {\it kernels\/}. Roughly speaking, the original image is converted to small ones with different features. For each kernel, a further processing called {\it max-pooling\/} is done on non-overlapping small patches, e.g., $2\times 2$ regions, that further coarse grain the arrays. Other pooling methods can be used. One can repeat such convolution+pooling layers a few times, but we will use one such combination layer in the percolation and two in the later part of the XY and the GXY  models. After this,  there is a fully connected layer of neurons, as in the FCN. This can be repeated a few more layers, but we will only use one such layer here. Finally, the fully connected layer is connected to a final output layer, and the number of neurons depends on the output type; for example, to distinguish digits 0 to 9, there are ten output neurons. For distinguishing between two phases, there are two output neurons.
Our optimization of the CNN again takes the advantage of the TensorFlow library.

\subsubsection{ Site and bond percolations}
\label{sec:cnnp}

\begin{figure}[thb]
\centering
\includegraphics[width=0.48\textwidth]{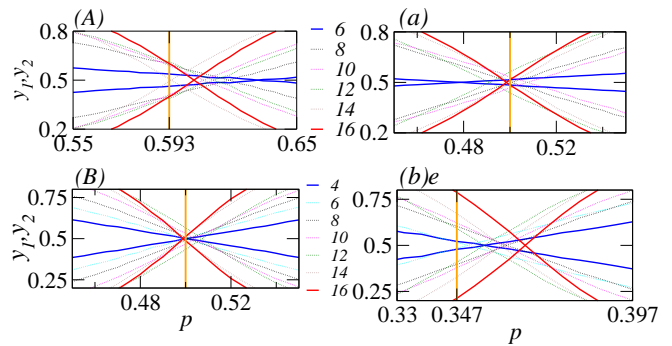}
\caption{(Color online)  The two values $y_1$ and $y_2$ of the output layer, using the CNN for site [(A)\&(a)]  and bond percolation [(B)\&(b)] on the square (left) and triangular (right) lattices, respectively.   }
\label{brc}
\end{figure}

We repeat the same study of percolation on square and triangular lattices with CNN. In the CNN structure,  we use  two combination layers (i.e., convolution+maxpool).
In Figs.~\ref{brc} (A) and (B), we show
the results of site percolation and bond percolation results on the square lattice, respectively. The critical point is converged to 0.593 and  0.5, respectively.
During the training, for the site percolation on e.g. the square lattice, the $L\times L$ site occupation configuration is used as input. For the  bond percolation, the $2L\times L$ bond configuration is used. As expected, the CNN works  well in learning the phase in percolation.
For the site and bond percolation on the triangular lattices, the results are also shown
in Figs.~\ref{brc} (a) and (b). We note that some slight improvement in the learning of the transition point can be made by careful choosing of $p$'s; see App.~\ref{sec:choosing}.
We also comment that since the Bethe lattices is not regular and we do not use CNN for the corresponding percolation problem. But it should be possible in principle. Moreover,
the FCN result of the percolation on the Bethe lattices is good enough.

\subsubsection{Labeling by cluster identifying algorithm}
\label{sec:lci}
\begin{figure}[htb]
\centering
\vskip 0.8 cm
 \includegraphics[width= 0.44\textwidth]{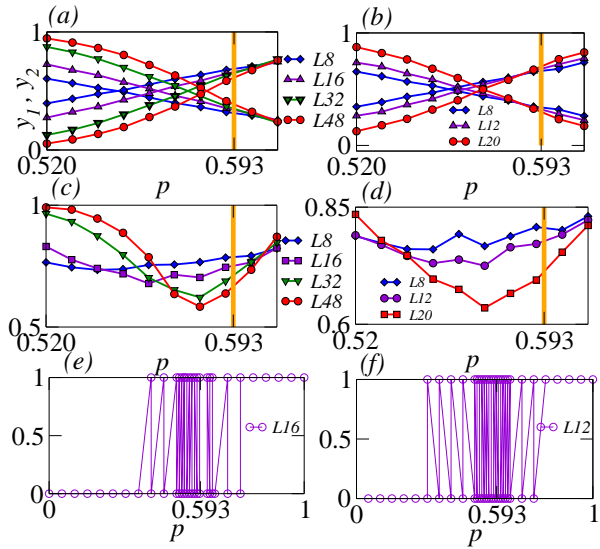}
 \caption{(Color online) The two values $y_1$ and $y_2$ of the output layer , using the cluster identifying method to label the configuration, with (a) the FCN and (b) the CNN.
       The respective accuracy curves of (a) and (b) are shown in  (c)  with FCN and in (d) with CNN.  Labels $\{0,1\}$ for percolation or not are shown for randomly generated configurations with  (e) $L=16$ and (f)  $L=12$. Notice the fluctuations near the transition.
 }
 \label{clabel}
 \end{figure}

In this subsection,  we would like to show
a different ways of training the data using the  non-local property of whether the configuration is percolating or not as the label ${\boldsymbol{y}}$.
In the previous works for thermodynamic phase transitions~\cite{juan1,un3, confu1},  for a configuration  $\boldsymbol{x}$ at  parameter such as $T$,
the corresponding label $y$ is set to be $0$ if $T<T_c$, which means that
the configuration is belonging to one phase. If $T>T_c$, the configurations
  belong to another phase, and the label is set
 be 1. Because
 the configurations around the percolation threshold  $p_c$ have  fluctuations and the configurations at $p<p_c$ may also  be percolating  and some configurations
  at $p>p_c$ may not be percolating, as illustrated in Fig.~\ref{clabel} (e) and (f).
Therefore, it is interesting to label the configuration  according to
whether or not the configuration has a spanning or wrapping cluster, instead of   the relationship between occupation probability $p$ and $p_c$.

Using the new labeling scheme, we show the results in Figs.~\ref{clabel} (a) and (c)  with sizes $L=8, 16, 32, 48$ by using FCN and  (b) and (d) with sizes $L=8, 12, 20$ by
using CNN.  Here no information about whether $p>p_c$ or $p<p_c$ is given to the network,
  and the labels of the  configurations  are obtained by
cluster-identifying algorithm~\cite{cluster-id}.
 However, the crossing obtained from the average values of the two output neurons gives the prediction of the percolation transition. They agree with the known results very well.

\section{The XY and GXY models}
\label{sec:XYmodels}
We now turn to the second type of models that we are interested in, which includes the pure XY model and the  GXY models.
As remarked in the Introduction, we shall first use  as input to the network
either projections of the  spin vector onto $x$ axis and $y$ axis, i.e.,
\begin{equation}
\boldsymbol{x} =(\text{cos}\theta_1, \text{sin}\theta_1, ..., \text{cos}\theta_N,\text{sin}\theta_N ),
	\label{eq:raw}
	   \end{equation}
	   or the spin orientations $\{\theta_i\}$.
The vorticity are defined as the winding numbers, i.e., a collection $\pm 1$ for
vortices and anti-vortices, but we shall not use those as input, for the reason remarked earlier due to the results in Ref.~\cite{xycnn}.
The results in Ref.~\cite{xycnn} show that
 the detection of
vortices does not necessarily result in the best classification accuracy, especially for lattices of less
than approximately 1000 spins. The advantage may show up for larger sizes, but the training becomes more costly. Here we limit ourselves to smaller sizes for using spin orientations (or their components) in the training, but later introduce a different approach in feature engineering that can be efficiently applied to larger systems.
In term of the two-dimensional image for the CNN, when we use the spin components, the $L\times L$ sites need to be effectively doubled to $L\times 2L$ that gives the same information of $\boldsymbol{x}$.

 In the following, we will  focus on the pure XY model on both the square and the honeycomb lattices and the  GXY model with $q=2, 3$ and $q=8$ on the square lattices.

\subsection{The pure XY model}
\label{sec:pxy}
\begin{figure}[h]
\centering
\includegraphics[width= 0.5\textwidth]{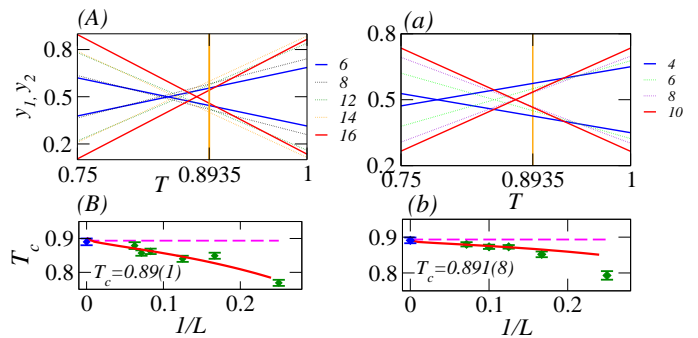}
\caption{ (Color online)
 The two values $y_1$ and $y_2$ of the output layer using the CNN for XY model by inputting
  (A)\{$\theta_i$\} with sizes $L=6,8,12,14,16$ and (a) \{sin $\theta_{i}$, cos $\theta_{i}$\}
   with sizes $L=4, 6, 8, 10$ on  the square    lattices.
(B) and  (b) Finite-size scaling of the critical points; the results obtained are consistent with $T_c= 0.8935$
within error bars for
both types of the inputs. }
\label{xy}
\end{figure}

\vskip 0.8 cm
\begin{figure}[hbt]
\centering
\includegraphics[width= 0.48\textwidth]{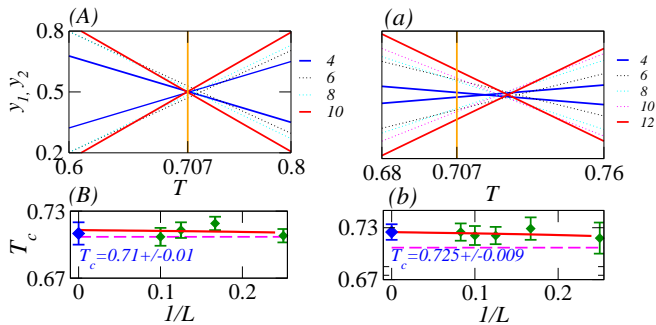}

\caption{ (Color online)
 The two values $y_1$ and $y_2$ of the output layer using the CNN for XY model by inputting
  (A)\{$\theta_i$\} with sizes $L=4, 6, 8, 10$ and (a) \{sin $\theta_{i}$, cos $\theta_{i}$\}
   with sizes $L=4, 6, 8, 10, 12$ on  the honeycomb    lattices.
(B) and  (b) Finite-size scaling of the critical points; the results obtained are consistent with $T_c=0.707$ using
both types of the inputs. }
\label{xytheta}
\end{figure}

The configurations of XY model on the square and honeycomb lattices are obtained by the classical Monte-Carlo
method; see e.g. Refs~\cite{sw,metro}.
    In the zero temperature limit, the spin at each site will point to the same orientation as the model is
ferromagnetic. However at a finite and small temperature $T$ less than $T_{KT}$, the spin orientations of spins points almost to the same direction with
some fluctuation, but there are excitations in the form of bound vortex-anitvortex pairs. Above the $T_{KT}$, the spin orientations  will become disordered as the vortex-anitvortex pairs unbind.

Since the phase transition points are already known for the XY model on  both lattices~\cite{xypoints, fssxy, o2}, i.e., $T_c^{\rm Square}=0.8935(1)$, and $ T_c^{\rm Honeycomb}=\sqrt{2}/2$, respectively,
it is thus interesting to see whether or not machine learning could recognize the phase transition
of XY model. Although the BKT transition of the XY model has been studied by machine learning
methods~\cite{xycnn}, it is our motivation to go beyond and study the GXY model.

In Figs.~\ref{xy}  (A) and (a), we show the learning  results  using the raw spin configurations  (i.e. both spin directions \{$\theta_i$\} and  spin components \{$\text{sin}\theta_{i}$, $\text{cos}\theta_{i}$\}) on the square  lattices with sizes $L=4$ to $L=16$.
The  lines $y_1$ \& $y_2$ correspond to the average values of the two output neurons.
Due to the fact that the performance of the network is lowest near the critical points,
   we use  25000 Monte-Carlo samples of the training data and of test data for each temperature $T$. In this way, we
  obtain results with low   standard deviations  around the critical point in the range $0.75<T<1$.

 Figures~\ref{xy}  (B) and (b) show the dependence on the lattice size $L$    of the estimated critical points $T_c$.   The green symbols are obtained from the  intersections.
{The red curves are the fitted  by the result from renormalization group ~\cite{tcxy}}:
\be
T_c(L)=T_c+\frac{b}{(\text{log}(L))^2},
\ee
where the  coefficient $b=\pi^2/4c$ and $c$ is a parameter.  In the thermodynamic limit, the estimated transition temperatures are  $T_c=0.89\pm 0.01$ and $T_c=0.891\pm 0.008$ for the pure XY model by using as the input  \{$\theta_i$\} and \{cos$\theta_i$, sin$\theta_i$\}, respectively. The results
 agree with  the   result of $T_c=0.8935$.
The CNN indeed works well in learning the BKT transition for the XY model, as previously demonstrated in  Ref.~\cite{xycnn} on the square lattice, so the success here comes with no surprise.

 Using the spin configurations on the honeycomb lattices as the input to the CNN gives good learning results  as shown in Fig.~\ref{xytheta}.
We note that the unit cell of the honeycomb
  has two sites.
  Each configuration thus have $N$ elements,
 where $N=2L\times L$ for using \{$\theta_i$\}  and $N=4L\times L$ for using \{cos$\theta_i$, sin$\theta_i$\}.
Our result agrees decently with the theoretical value  $T_c=\sqrt{2}/2\approx 0.707$. (Some slight improvement can be made by choosing training data generated at temperatures symmetric about $T_c$ and without including those at $T_c$; see App.~\ref{sec:choosing} and Fig.~\ref{fig:imp}.) In the next section we will study the  GXY model to extend the machine learning  beyond the XY model.
\subsection{The generalized  XY models}
\subsubsection{q=2 and q=3}
\label{sec:pgxy}

\begin{figure}[htb]
\vskip 0.5 cm
\centering
\includegraphics[width=0.42\textwidth]{Fig12a.eps}
\includegraphics[width=0.42\textwidth]{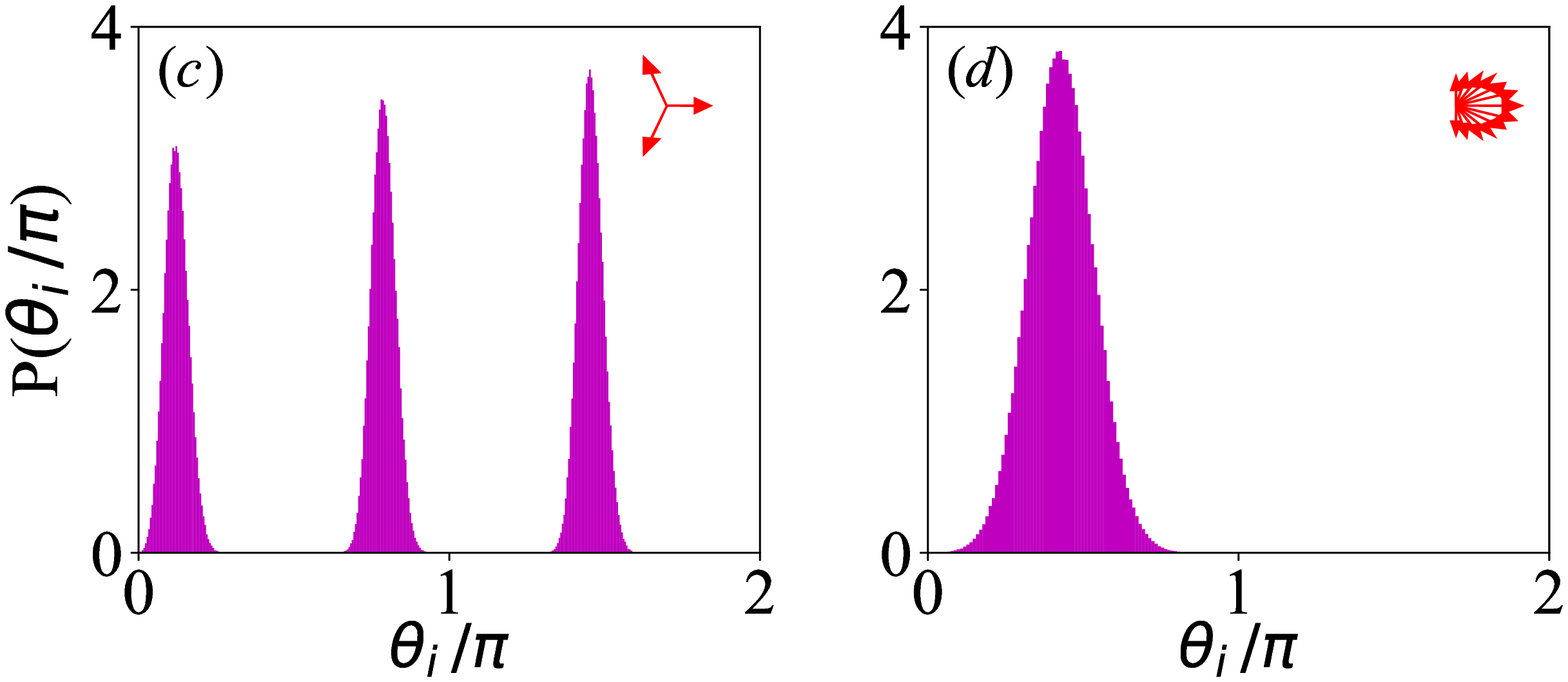}
\includegraphics[width=0.42\textwidth]{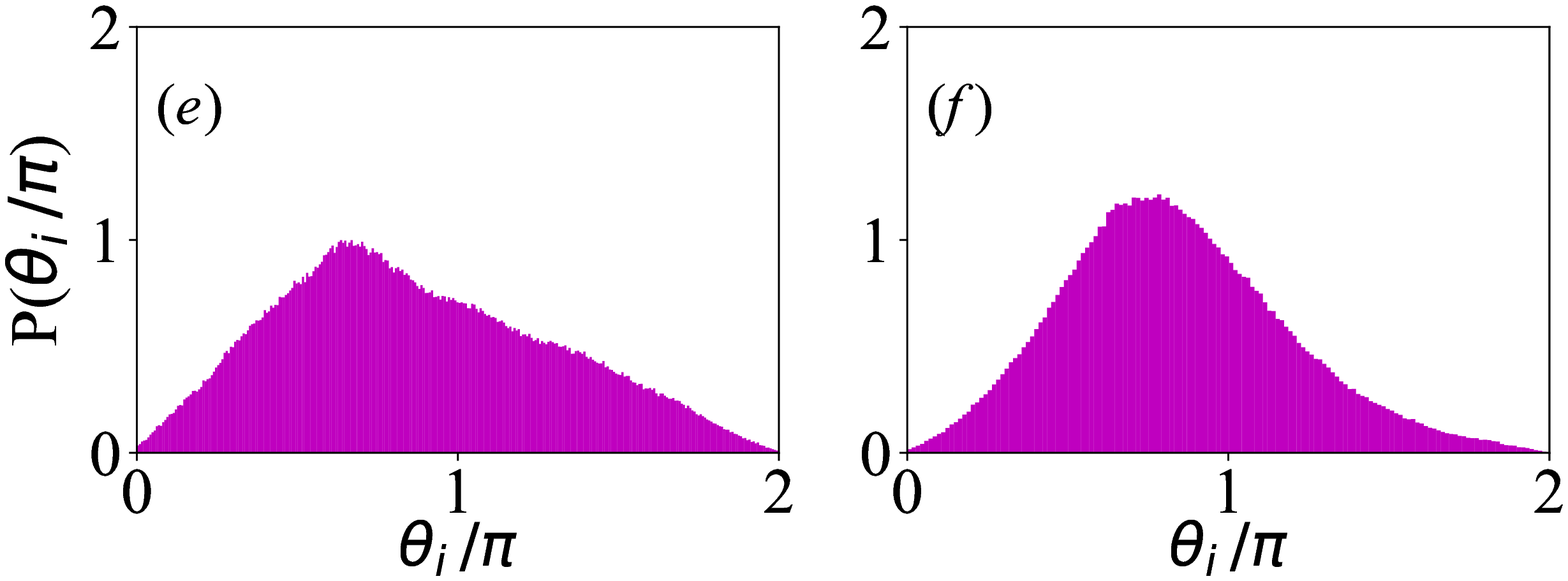}
\caption{Phase diagrams of the GXY model for (a) $q=2$ and (b) $q=3$, the data are from Refs.~\cite{xyq2,xyq3}. The symbols  N, P and F represent the nematic, ferromagnetic and paramagnetic phases, respectively. The dashed lines show the parameter paths to be scanned.	 (c) Histograms of the configurations $\theta_{i,j}$ for
the nematic phase at  $q=3$, $\Delta=0$ and $T=0.2$,
the three independent peaks show three preferred  orientations.
(d) Histogram for the spin orientations in the (quasi-long-range) ferromagnetic phase at $q=3$, $\Delta=1$, $T=0.2$, where the system prefers one spin angles with some fluctuations. {(e) \& (f) Histogram for  the paramagnetic phase at $q=2$, $\Delta=0$, $T=2$ (e), and the paramagnetic phase at $q=3$, $\Delta=0$, $T=2$ (f), respectively. Note that for illustrations, these distributions are obtained by  over 2000 samples. But for the input to the neural network, we  use histograms each derived from a very small number of samples, such as 20.}}
\label{qhist}
\end{figure}

\begin{figure}[htb]
\centering
\vskip  0.5 cm
\includegraphics[width=0.48\textwidth]{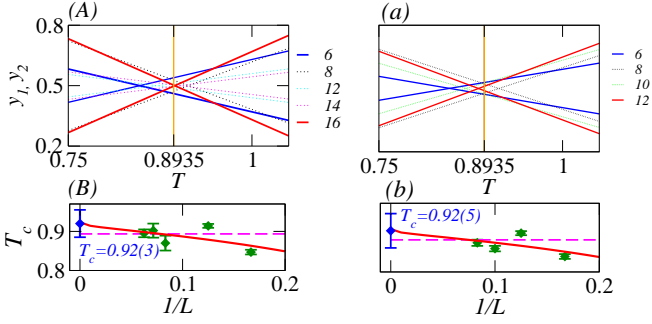}

\caption{(Color online)
The two values $y_1$ and $y_2$ of the output layer of learning of
(A) \{$\theta_{i}$\}  and (a) \{sin $\theta_{i}$, cos $\theta_{i}$\}  for (a) $q=2$, $\Delta=0$    with
various lattice sizes.
(B) and (b) The estimated critical points from the dependence on lattices sizes $1/L$.
In the thermodynamical limit, $T_c=0.92(3)$ and $T_c=0.92(5)$ are estimated. }
\label{sinq2q3}
\end{figure}
\begin{figure}[htb]
\centering
\vskip  0.5 cm
\includegraphics[width=0.48\textwidth]{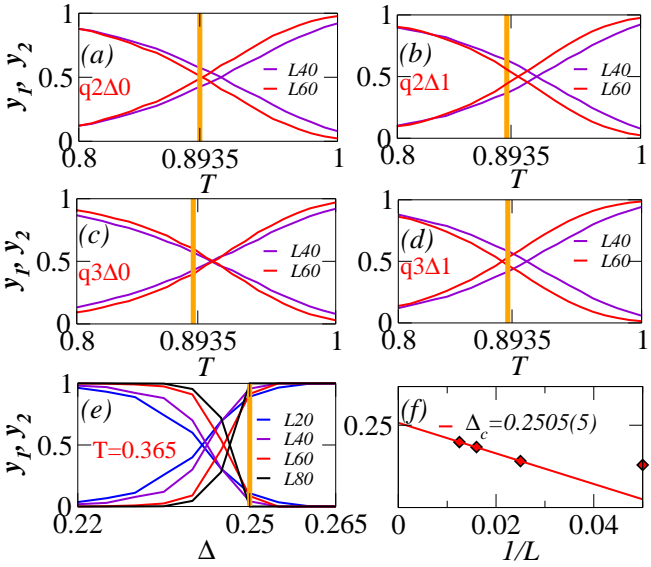}
\caption{(Color online) The two values $y_1$ and $y_2$ of the output layer in the learning of  histograms of \{$\theta_{i}$\} for (a) $q=2$, $\Delta=0$; (b) $q=2$, $\Delta=1$;
 (c) $q=3$, $\Delta=0$; (d) $q=3$, $\Delta=1$.
(e) The average results in the  output layer and (f) the finite-size scaling of the critical points  by scanning the parameter $\Delta$, and critical point $\Delta_c$
is estimated to be $0.2505(5)$.}
\label{q23}
\end{figure}
The phase diagram of    GXY models \cite{xyq2,xyq3} is rich and two examples with $q=2$ and $q=3$ are shown, respectively, in Figs.~\ref{qhist} (a) and (b).
In the lower temperature and $\Delta=0$ limit,
the system is in the generalized nematic phase that   has $q$
preferred spin orientations. The statistical distribution for spin orientations in
the  nematic phase displays $q$ peaks as shown in Fig.~\ref{qhist}(c).
In the $\Delta=1$ limit, the system is quasi-long-range ferromagnetic (broken reflection symmetry) in low-temperature limit and the distribution of the spin orientations is shown in
Fig.~\ref{qhist}(d).
In the higher temperature, the system become disordered, and is in a paramagnetic phase.    The distribution for the paramagnetic phase are also shown in
Figs.~\ref{qhist}(e) and (f) for $q=2$ and 3, respectively. Clearly, the distributions spread through a very wide arrange of angles due to strong thermal fluctuations.

We use the configuration of (1) \{$\theta_{i} $\} and (2) \{$\cos\theta_{i},\cos\theta_{i}$\} as input to the CNN  with two convolutional layers, and  the network  indeed can learn the transition.
For both $\Delta=1$ (pure XY model) and  0 (the GXY model)    the phase transition is located at $T/J=0.8935$.
As remarked earlier,  this is because the $\Delta=0$ GXY model isomorphic to the usual XY model~\cite{rep1} by changing the variable $q\theta_i$ as $\bar{\theta_i}$.
For  both $\Delta=0$ and $\Delta=1$,  using either (1) or (2) as the input
works well and the CNN can distinguish, respectively, the quasi-long-range ferromagnetic phase and the nematic phase,  from the  high temperature disordered paramagnetic phase.
Fig.~\ref{sinq2q3} (A)  displays the average values of the output layer and the performance of learning by using
\{ $\theta_{i}$ \} as input  for  $q=2$ at $\Delta=0$,  with $L=6,8,12,14, 16$, while that in (a)
is obtained from using  \{sin $\theta_{i}$, cos$\theta_{i}$\} as input.  The resulting two curves $y_1$ and $y_2$ from the two neurons in the output layer
can distinguish different phases and their crossing gives the transition.
In the thermodynamic limit, the critical points are estimated to be $T_c=0.92\pm 0.03$ and $T_c=0.92\pm 0.05$, respectively, using the two different types of input.

Here in the network there are 32 and 64 kernels in the first and second layers, respectively.
We use the Wolff-cluster algorithm to generate  configurations  for the GXY model.  To obtain enough equilibrated states, we throw out the configurations during the  first  10000 Monte-Carlo steps.
To avoid the correlations between configurations, we pick up configurations with interval of 2-5 Monte Carlo steps at each temperature.

 Inspired by the main difference between the above three phases being the shape of the histograms,
 we investigate whether using such feature engineering (i.e. histograms) can help the learning better.
In principle, spin configurations from each sample in the Monte Carlo algorithm generates a histogram.
However, to make the
histogram smooth, we use multiple configurations to average (e.g. 20) for small system sizes.
Employing wolff cluster algorithm allows us to access larger lattices, we can use just one single configuration to generate a histogram.  After obtaining the histogram, we
segment the images into a $32 \times 32 $ matrix of black and white pixels, in which
the white area is set to  0 and colorful area is set to  1.
These matrices of pixels are our engineered feature and are used as the input to the CNN for training.

We directly recognize the histograms and obtain results as accurate as other kinds of inputs, using the two-layer-convolution CNN as shown in Figs.~\ref{q23} (a)-(d) along the four red dashed lines in Figs.~\ref{qhist} (a)-(b). For both $\Delta=0$ and  1, the GXY model has  the phase transition located at $T/J=0.8935$.  The results of using histograms as the engineered feature make the CNN able to recognize different phases in the generalized XY model and the associated transitions in these two limits of $\Delta$.  For completeness, we also scan a path in the phase diagram of $q=3$ model by first varying $\Delta$  at $T/J=0.365$ (for which the $\Delta_c=0.25$~\cite{xyq3}), and the results   shown in Figs.~\ref{q23} (e)-(f) demonstrate that  the neural network, in particular, can also distinguish  the
nematic phase from the ferromagnetic phase and learn the transition point.
The approach of using histograms helps us to access larger lattices without any further training.

\begin{figure}[ht]
\centering
\includegraphics[width=0.4\textwidth]{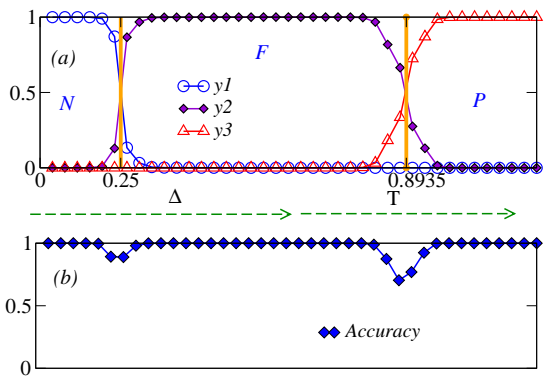}
\caption{(a) The results in averaging outcomes separately in the  three neurons of the output layer by  scanning the green dashed lines in Fig.~\ref{qhist}(b).
(b)  The accuracy of the learning. }	
\label{3phases}
\end{figure}

Armed with the success of learning two distinct phases,  we move on to test whether our method can be used to distinguish three phases. In particular, we scan the phase diagram in Fig.~\ref{qhist} (b) along the
green dashed lines, combining the previous horizontal path varying $\Delta$ (at $T/J=0.365$) and then the vertical path by varying $T/J$ (at $\Delta=1$); this path cuts through the three phases: N, F and P.
To do this, we need to use three neurons in the output layer. During the training of the network,  the configurations in the three phases are labeled as  0, 1 and 2 respectively. The sigmoid function maps
the output layer located  between the range $[0, 1]$.
In the test stage, the first neuron in the output layer  (representing the N phase) is 1 for $\Delta<0.25$ and becomes zero at other  two
phases (the F and P phases).
The  output of the other two  neurons shows  converse behavior as shown in Fig.~\ref{3phases} (a). There are  three curves corresponding to the three neurons in the output layer.  The accuracy is shown in Fig.~\ref{3phases} (b) and it equals to 100\% at non-critical regimes and decreases near the two
critical points around $(\Delta=0.25,T=0.365)$ and $(\Delta=1,T=0.89)$. In short,  we have demonstrated successful learning of  three phases (N, P
and F) and the transitions.

\begin{figure}[hbt]
\centering
\includegraphics[width=0.22\textwidth,height=0.15\textwidth]{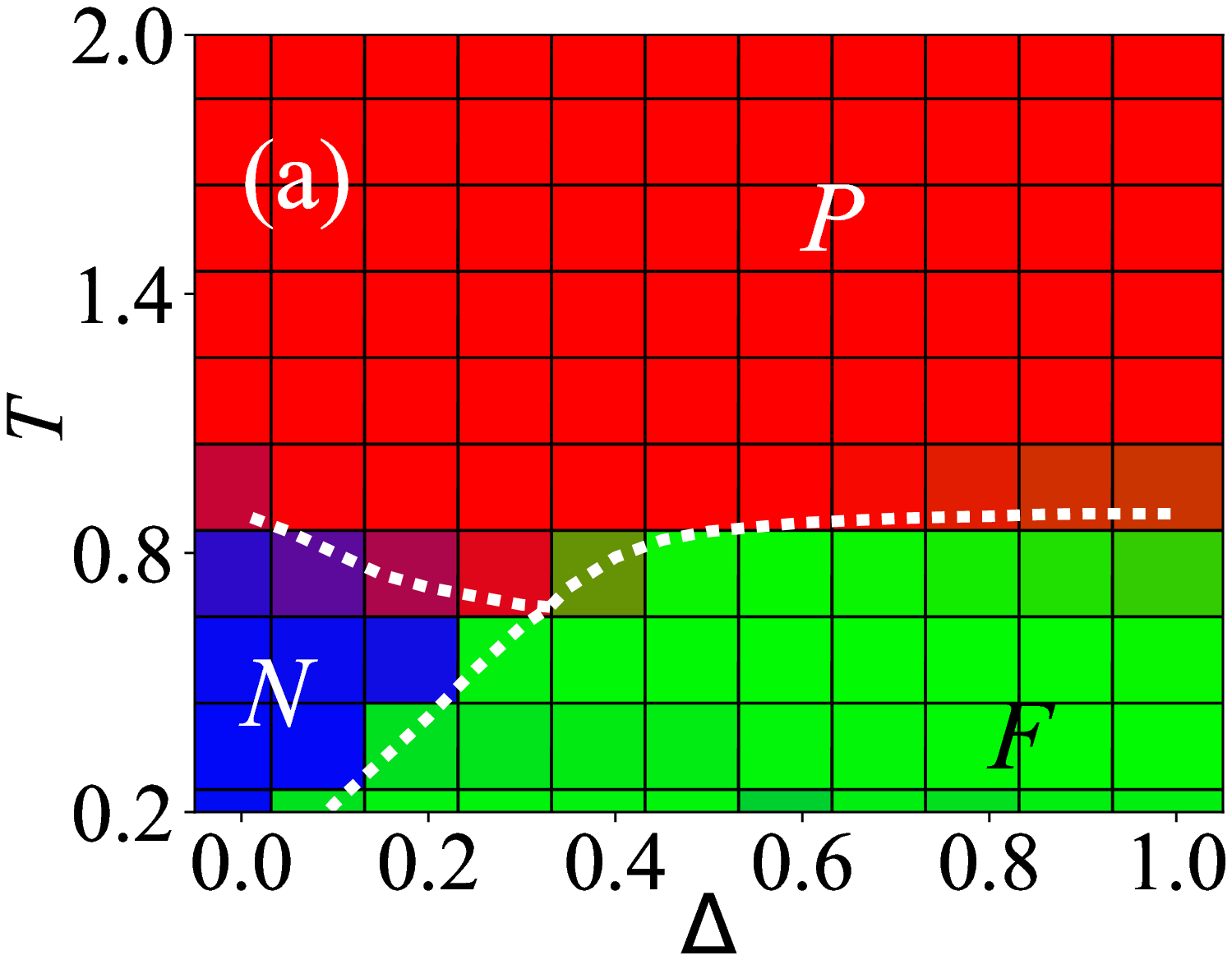}
\includegraphics[width=0.22\textwidth,height=0.15\textwidth]{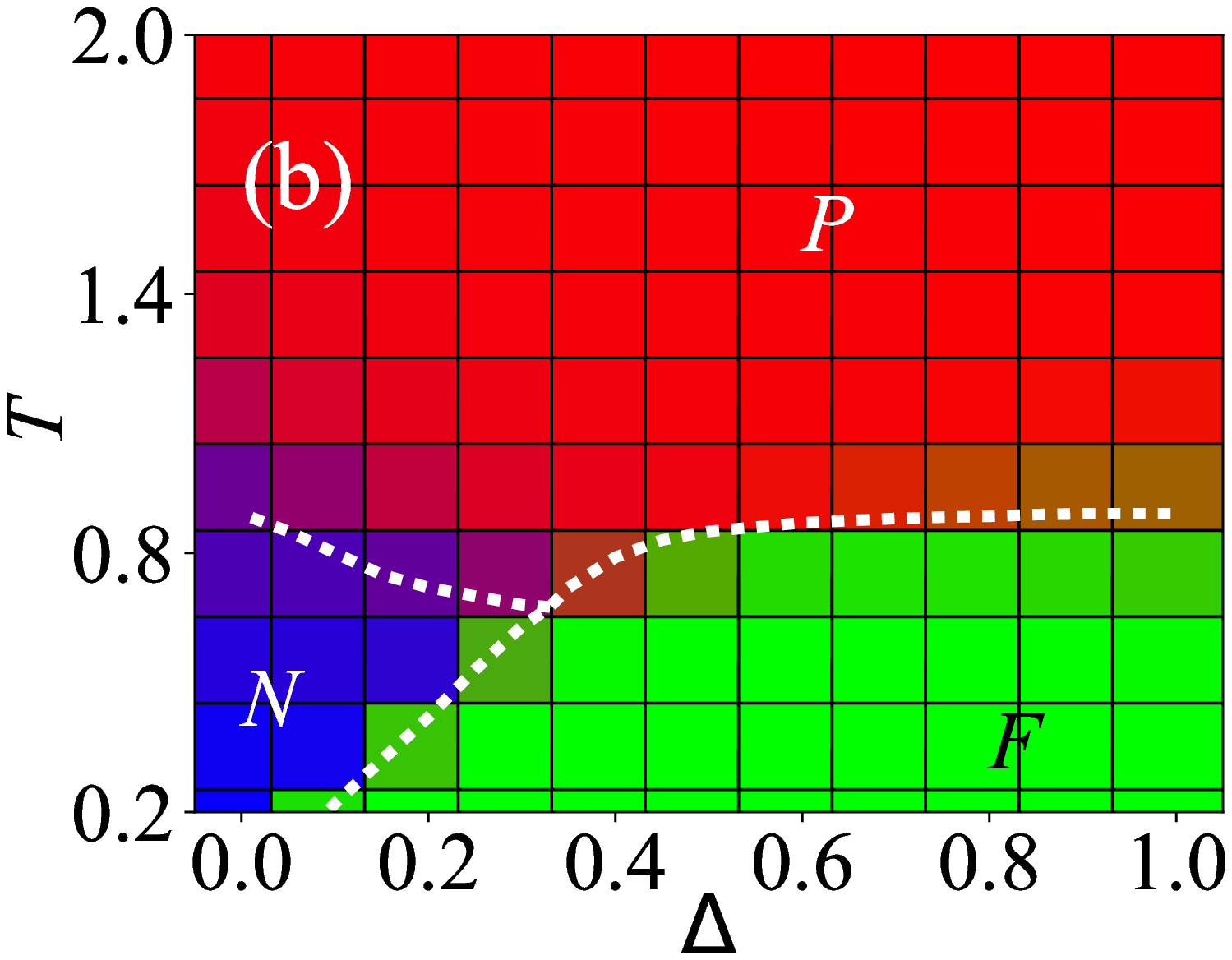}
\includegraphics[width=0.22\textwidth,height=0.15\textwidth]{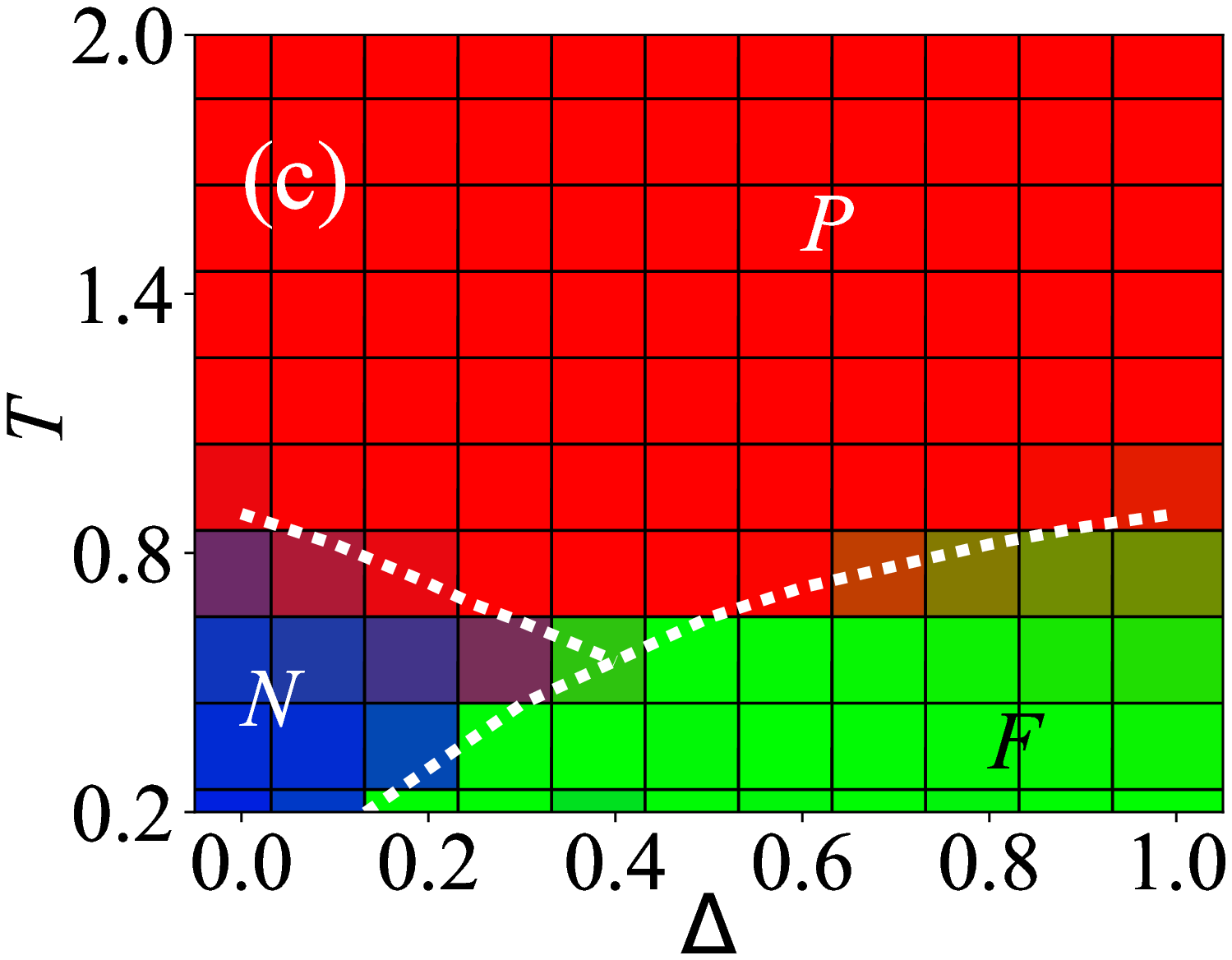}
\includegraphics[width=0.22\textwidth,height=0.15\textwidth]{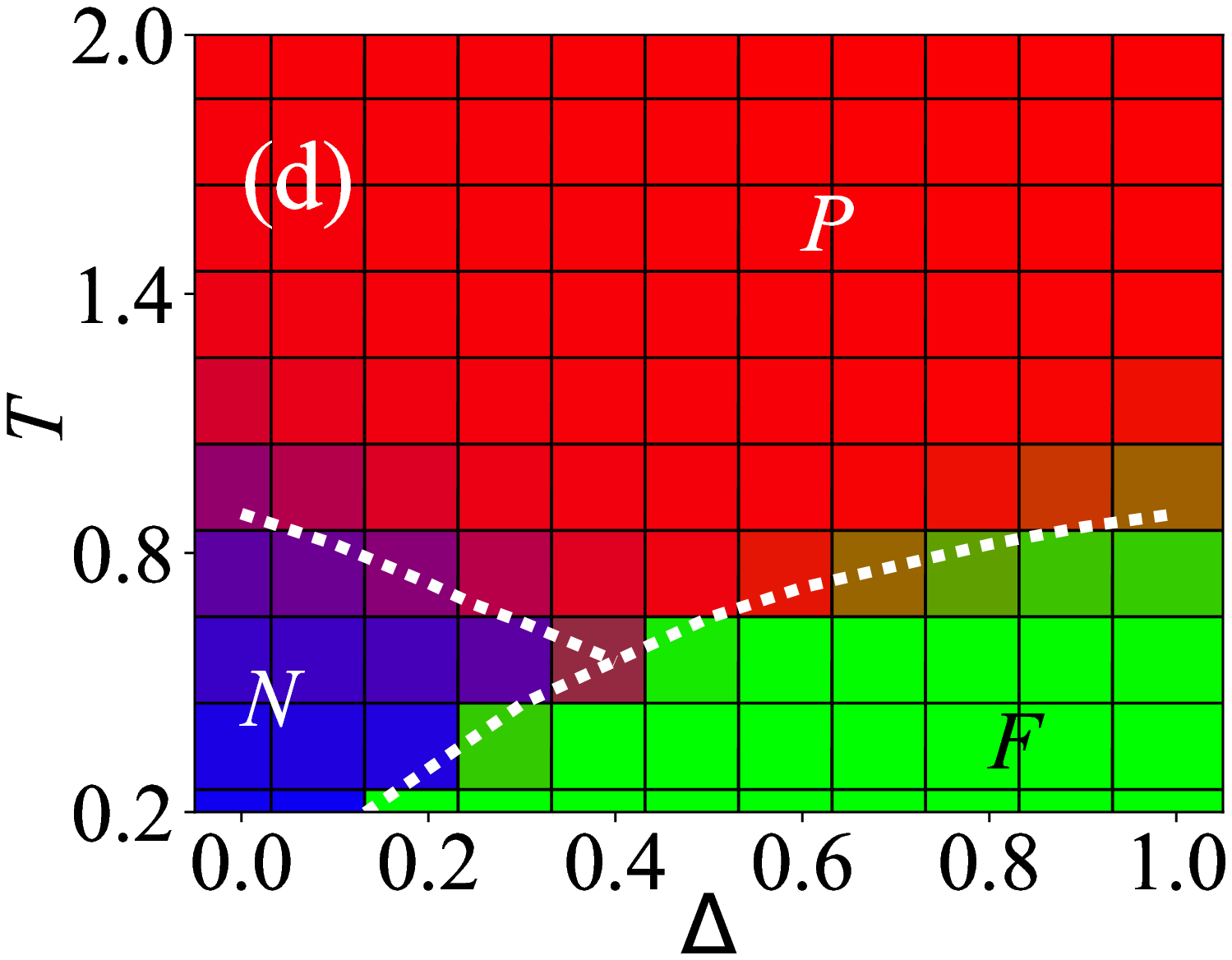}
\caption{The phase diagram of the  GXY model, which contains, Nematic phase (N, blue), Ferromagnetic phase (F, green), and paramagnetic phase (P, red),  obtained by the semi-supervised method discussed in the text. The neural network is trained only for data from $\Delta=0$ and $\Delta=1$.(a) The $q=2$ model: the input data for training is the histogram of spin orientations  \{
$\theta _i$ \}.  (b) The $q=2$ model: the input data for training is the  spin orientations. (c) The $q=3$ model:  the input data for training is the histogram of spin orientations. (d) The $q=3$ model: the input data for training is spin orientations {$\theta_i$}. These agree with those (from Monte Carlo simulations) in Figs.~\ref{qhist}(a)\&(b).   The phase boundaries represented by white lines by Monte Carlo methods are also shown.}						                      \label{semisuper}
\end{figure}
													
As a final test, we use a semi-supervised method to retrieve  the global phase diagram of  GXY model
with $q=2$ and $q=3$ on the square lattice of $12\times 12$ sites.  What we mean by the  semi-supervised method is that,  only the limited data  that are not generic are used in the training, for example, those along $\Delta=0$ and $\Delta=1$, i.e., the two vertical paths  in Fig.~\ref{q23}(a) and (b). These are non-generic, as they only give very limited representations in the phase diagram. Once the network is trained and optimized, we use the neural network to predict the phases, using the configurations generated from Monte Carlo in the whole phase diagram, with parameters  ($\Delta$, $T$) covering the range $\Delta=0, 0.1, 0.2,\cdots, 1$ and $T=0.2, 0.4,\cdots, 1.8,2$. We use both the spin orientation and the histogram   as the input, and
the phase diagrams thus obtained, as shown in Fig.~\ref{semisuper},  agrees  well with those in  Figs.~\ref{qhist} (a)\&(b).
 \subsubsection{q=8}
 For $q>3$, the results in Refs~\cite{ gxyq8, gxy14}  suggest the existence of new phases at
intermediate values of $\Delta$ that do not appear for either $\Delta=0$ or 1. Since
the existence and/or nature of some of those transitions are still disputed, it
would be interesting to see the outcome of the neural networks in those cases.

\begin{figure}[t]
\centering
\includegraphics[width=0.25\textwidth]{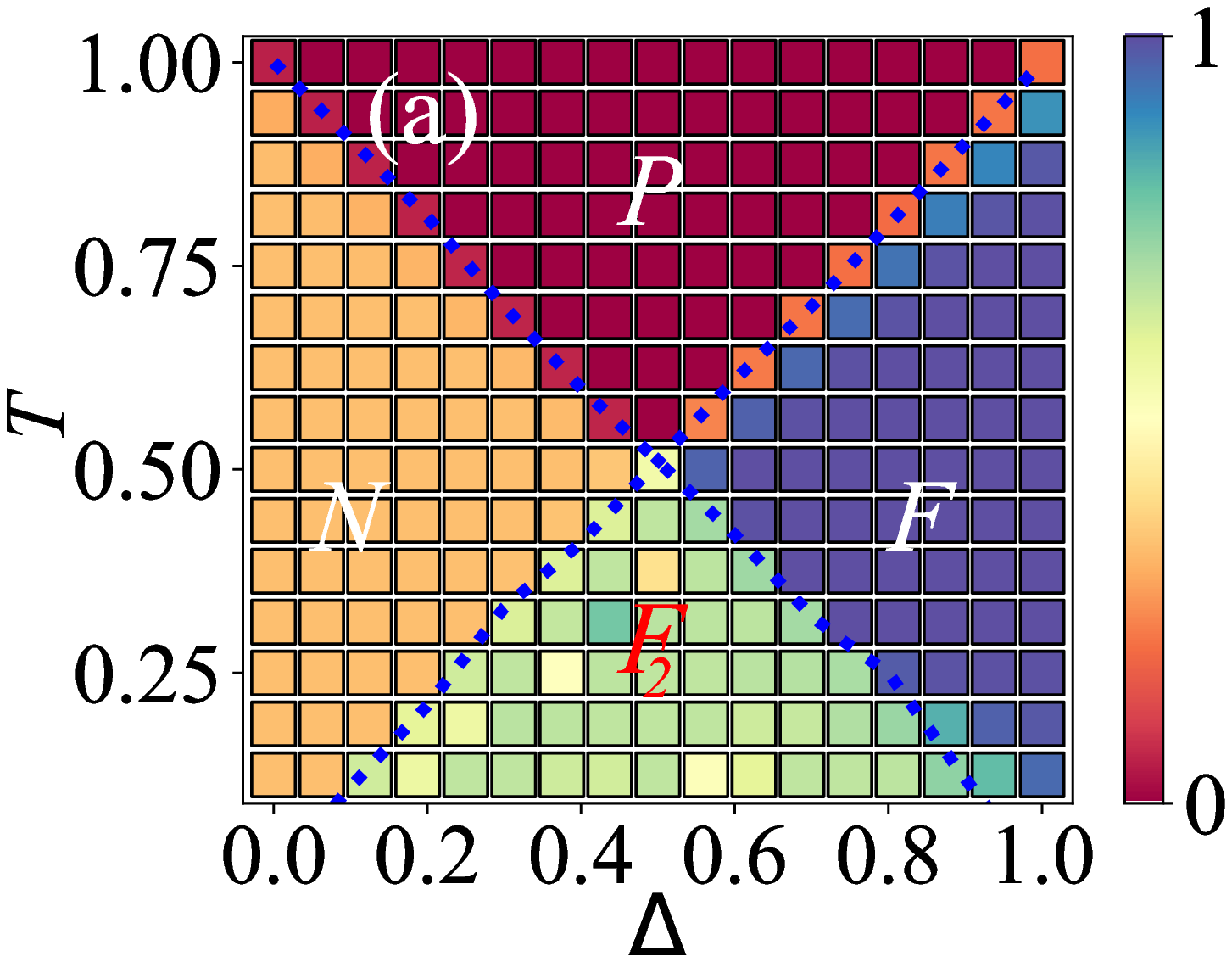}
\includegraphics[width=0.20\textwidth]{Fig17q8bcd.eps}
\includegraphics[width=0.225\textwidth]{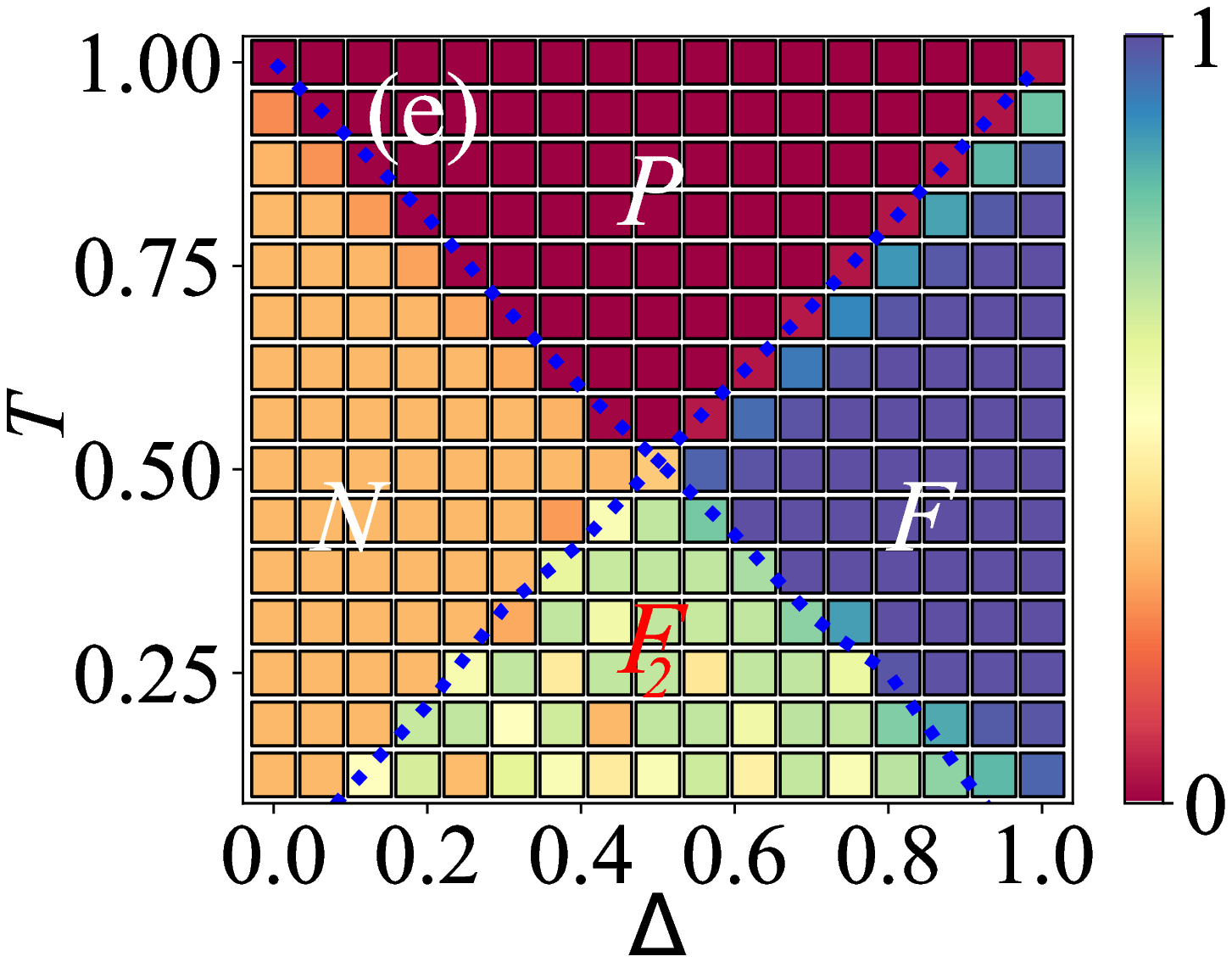}
\includegraphics[width=0.225\textwidth]{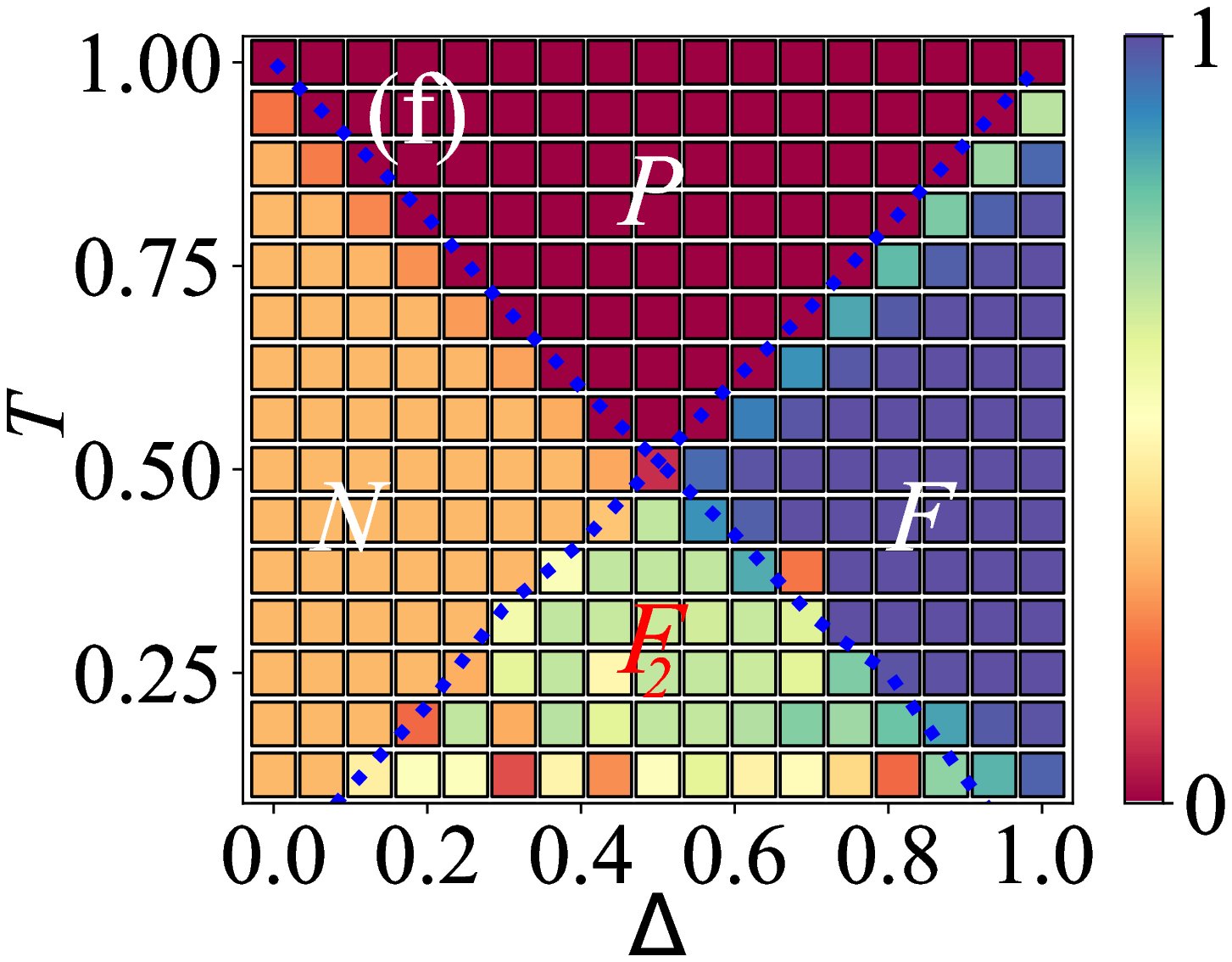}
\caption{{(a) The phase diagram of the GXY model with $q=8$, containing N, P,  $F$, and  $F_2$ phases from Ref.~\cite{gxyq8}. The network is trained from the spin histograms of size $L=100$. (b) Illustration of the distributions of spins directions  in the nematic phase show 8 peaks.
 (c) The distributions of spins directions in  $F$ phase. Using the trained network of (a), we obtain
 the  phase diagrams of systems with sizes $L=150$ (e) and $L=200$ (f).}}
 \label{fig:phq8}
\end{figure}

Without loss of generality,
  a typical value  $q=8$ is chosen.
In Fig.~\ref{fig:phq8} (a), the phase diagram, containing N, P,  and F phases, is shown.
The blue dashed lines are the data from Refs~\cite{gxyq8, gxy14}. The color of the small  squares  is the mapping of the average values in the output layer $y_N$, $y_{F_2}$, $y_P$, and $y_F$ of the two-layer CNN.
The color is mapped via $z=y_P+ 20*y_N+ 40*y_{F_2} +60*y_F$ and  its normalized expression $z=[z-min(z)]/max(z)$.
Besides the previous phases of the $q=3$ model, a new phase  $F_2$ phases emerges, consistent with Refs.~\cite{gxyq8, gxy14}.

Here we comment on
the way we train the CNN.
Firstly, the lattice size should be large enough such as $L=100$.
By fixing the temperatures at  $T=0.35$ and $0.8$, we scan the parameter $\Delta$ from 0 to 1
in the phase diagram with an interval of 0.05. For each parameter point,  2500 histogram samples for training  are used and  25000 samples of histograms for testing, and  four labels for the different phases are used. We can distinguish the new emerged phase $F_2$ by the non-zero labels. At the same time, the
 boundaries between other phases are also obtained, consistent with the known results (indicated  by the blue dotted lines).

In Figs.~\ref{fig:phq8} (b)-(d), the distributions of spins directions $P(\theta/\pi)$ for each phase are shown according to configurations, generated by the Wolff-cluster Monte-Carlo method.
Different from the case of $q=3$,  $P(\theta/\pi)$ in the $q=8$ nematic phase has 8 peaks
 at $T=0.35$, $\Delta=0$.  In the $F_2$ phase, as shown Fig.\ref{fig:phq8} (c), some spin directions (peaks)  dominate the configuration at  $T=0.35$, $\Delta=0.5$. In
the $F$ phase, all of the directions for the spins are restricted in   a half  plane $0<\theta<\pi$.

  In order to have efficient machine learning algorithms with
 large data sets, the representation of the data (e.g. configurations
from Monte Carlo simulations) can affect both efficiency and accuracy.
 In this paper, one new idea is to use distribution of spin orientations
as the input to the neural network, instead of the complete spin
 configurations. In the usual machine learning, the size of the latter ($L\times L$ spins) can be so large that the training time can take too long. However, transforming
the data into distributions can reduce the cost of training, as the
distribution of the data is a much smaller set.

Furthermore, the neural  network trained by histograms of spin orientations from a system of size $L\times L$ can be used to test the distributions
from other system sizes without further training.
In Figs.~\ref{fig:phq8} (e)-(f), the phase diagrams, consistent with the previous one, are obtained  by inputting the test data from systems with sizes $150\times 150$ and $200 \times 200 $, respectively. The  network used for the test is  from that training using data from the $100\times 100$ system size.
We note that the size of our histogram images was chosen and fixed more or less arbitrarily at, e.g., $32 \times 32$, and that other (larger) sizes can be used to increase accuracy. Using  previously trained network
avoids training the network repeatedly for other system sizes.



\section{Conclusion}
\label{sec:con}
In summary, we  have used machine learning methods to study the percolation,  the XY model and the  GXY model  in the two dimensional lattices. For the percolation phase transition, the unsupervised  t-SNE can map the high dimensional data-sets of configurations into an two-dimensional image with classifiable data. Using the FCN even without explicitly giving the two-dimensional spatial structure, still allows to  recognize  the percolation phase transition.  By feeding the information about the existence or not of the spanning cluster, the transition can be predicted without any training information on whether the configurations are generated with $p>p_c$ or $p<p_c$.  The percolation exponent $\nu$ was obtained correctly using results from the output neurons.
We have also demonstrated that the CNN method works well for percolation, but there is no substantial advantage using CNN. The advantage of the CNN against the FCN arises in the study of the XY model and the generalized XY models.

The pure XY model on the square and honeycomb lattice in our study was learned by
inputting the spin configurations \{cos$\theta_i$, sin$\theta_i$\},\{$\theta_{i,j}$\} and even with small sizes such as $L=4, \cdots, 16$, the critical point could be obtained by performing  the finite-size scaling.
For the generalized XY model with $q=2$, 3 and 8, the global phase diagrams were obtained by a semi-supervised method, i.e., with a network trained by  learning just some limited set of the data.
Specifically, for $q=8$, the new phase $F_2$ in the range of  $0 <\Delta < 1$   is also be confirmed through the perspective of
machine learning.

The use of spin configurations as the naive input  works for training the network to recognize phases in the XY model. One key difference between the phases is the probability distribution of the spin orientations.
We have devised a feature engineering using  the histograms  of the spin orientations instead, and this has resulted in successful learning of various phases in the generalized XY model beyond the XY model.
 Moreover, the trained
network  with system size $L\times L$ can also be used for testing data from other system sizes
($L'\times L'$, where $L'\ne L$), saving additional training effort.
The use of machine learning in phases of matter in general may still need the ingenuity of appropriate features for the neural network to learn, but can become a useful tool.

\appendix
\section{t-SNE method}
Here we summarize for convenience the t-Distributed Stochastic Neighbor Embedding (t-SNE) method~\cite{tsne1}, which is an improvement from the Stochastic Neibhgor Embedding (SNE) method~\cite{sne}. The main idea of these methods is to, from a set of high-dimensional data, represented by high-dimensional vectors $\{x_i\}$, obtain a corresponding set of low-dimensional (e.g. 2- or 3-dimensional) data, represented by a set of vectors $\{y_i\}$ such that the latter maintains key features of the former. In the t-SNE method, the pairwise similarity $q_{ij}$'s for a pair of low-dimensional vectors $y_i$ and $y_j$ is defined as
\begin{equation}
     q_{i,j}=\frac{(1+||y_{i}-y_{j}||^{2})^{-1}}{\sum_{k \ne l}(1+||y_k-y_l||^{2})^{-1}},
\end{equation}
whereas that for the high-dimensional vectors is defined as $p_{i,j}\equiv(p_{j|i}+p_{i|j})/(2n)$, where $n$ is the total number of vectors and
\begin{equation}
     p_{j|i}=\frac{\exp(-||x_{i}-x_{j}||^{2}/2{\sigma}_i^{2})}{\sum_{k \ne l}{\exp(-||x_{k}-x_{l}||^{2}/2{\sigma_i}^{2})}}.
\end{equation}
The lower dimensional vectors $y$'s are obtained by using a gradient descent approach by minimizing the cost function,
\begin{equation}
    C={\sum_{i}\sum_{j}p_{ij}\log\frac{p_{ij}}{q_{ij}}},
\end{equation}
where the gradient by varying $y_i$'s is given by
\begin{equation}
   \frac{{\delta}C}{{\delta}y_{i}}=4\sum_{j}(p_{ij}-q_{ij})(y_{i}-y_{j})(1+||y_{i}-y_{j}||^{2})^{-1}.
\end{equation}
In the above, the standard deviation $\sigma_i$'s are obtained by fixing the so-called perplexity (supplied by the user),
\begin{equation}
{\rm Perp}(P_i) = 2^{-\sum_j p_{j|i}\log_2 p_{j|i}},
\end{equation}
which is typically chosen between 5 and 50, according to the performance. Here we set it to be 20.

The initial low-dimensional vectors $y_i$'s are generated randomly. Iterating the gradient descent procedure will yield an improved approximation consecutively, until the gradient is very small.
\section{activation functions of sigmoid and softmax}

\begin{figure}[hbt]
\centering
\includegraphics[width=0.45\textwidth]{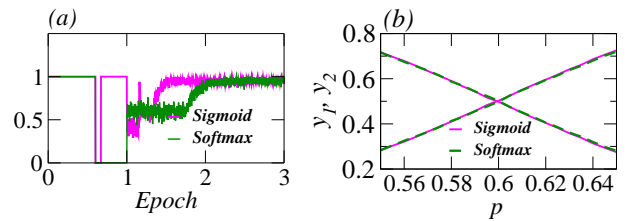}
\caption{(a)The training accuracy by using the  sigmoid and softmax activation, respectively.
(b) For the site percolation on the square lattices, we find the average output layers are the same using enough samples (25000).  }
 \label{fig:sig}
\end{figure}

There are many activation functions to be used in the neural network, here we use the sigmoid function for the FCN.  Taking
  the site percolation as an example for test,
  we output  the results with 25000 samples per occupation probability.
  In Fig.~\ref{fig:sig} (a),
 the training accuracy of the using sigmoid function  reaches an equilibrated  stage faster than
that by the softmax function.
  It is also found that the results of using  softmax and sigmoid function  are almost the same as shown in Fig.~\ref{fig:sig} (b).
 Therefore, enough samples can avoid
systematic error induced by different activation functions.

\section{Choosing training data}
\label{sec:choosing}
\vskip 1 cm
\begin{figure}[hbt]
\centering
\includegraphics[width=0.45\textwidth]{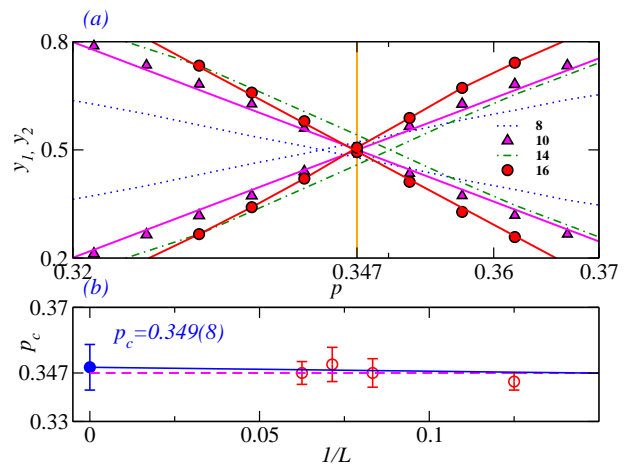}
\caption{The result of bond percolation on the triangular lattice via training the CNN with configurations generated symmetrically w.r.t. $p_c$. This is to be compared with Fig.~\ref{brc}(b), and the results here show some improvement.}
 \label{fig:pcwn}
\end{figure}

There are several factors that could affect the trained network.
In this appendix,  we compare   two  choices of the training data and the resultant learned transition points, i.e. intersections of $y_1$ and $y_2$. The first approach for the training data is  to use configurations corresponding to probabilities $p$'s uniformly, as we have used in most of the plots in the main text, such as the percolation in Fig.~\ref{brc}. The second choice is to use configurations generated at $p$'s symmetric with respect to $p_c$ (but not including those at $p_c$), which is demonstrated at Fig.~\ref{fig:pcwn}.

We remark that the training data used in Fig.~\ref{brc} were obtained at $p$'s that are uniform. By judiciously making the choice such that these $p$'s are symmetric with respect to the transition point (excluding the transition point), better learning of the transition can be made.

\begin{figure}[hbt]
\centering
\includegraphics[width=0.45\textwidth]{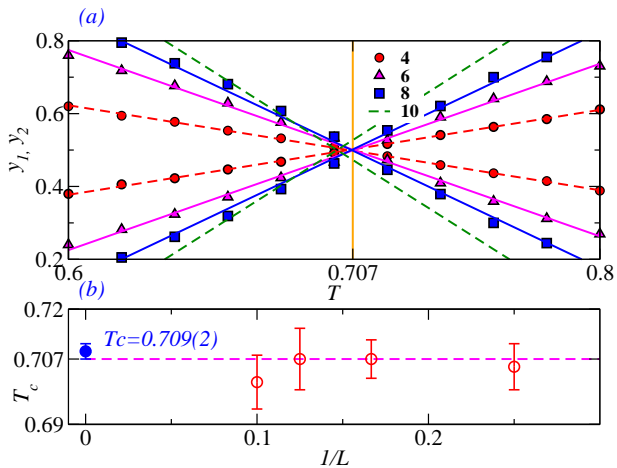}
\caption{(a) Averaged values of the two neurons in the output layer, $y_1$ and $y_2$, for the XY model on the honeycomb lattice by
choosing the training data $\{cos\theta_i, sin\theta_i\}$ below $T_c$ and above $T_c$ symmetrically with respect to $T_c$ but not including it. (b) The finite-size scaling of $T_c$ and  the value in the thermodynamic limit
is estimated to be 0.709(2). }
 \label{fig:imp}
\end{figure}

We also test this for the XY model on the honeycomb.
In Fig.~\ref{fig:imp} (a), $y_1$ and $y_2$ for the two neurons in the output layer are plotted and  the result $T_c=0.709(2)$ from finite-size scaling in Fig.~\ref{fig:imp} (b)   appears more accurate than $T_c=0.729(8)$ in Fig.~\ref{xytheta} (b).

\vskip 0.5 cm
\begin{acknowledgments}
WZ thanks for the valuable discussion with C. X.  Ding on Monte-Carlo simulations and  gratefully acknowledge financial support from China Scholarship Council and the NSFC under Grant No.11305113; TCW is supported by the
National Science Foundation under Grants No. PHY 1314748 and No. PHY
1620252.
\end{acknowledgments}

\noindent {\it Notes added\/}. After we finished the manuscript, we learned a very interesting paper on ``Parameter diagnostics of phases and phase transition learning by neural networks'' by Suchsland and Wessel~\cite{Wessel}. Even though it is beyond the scope of the current manuscript, but as a future direction, it will be interesting to employ the methods presented there in percolation and  the generalized XY models. (They have already analyzed the XY model.)

\end{document}